\documentclass[superscriptaddress, showpacs, floatfix, twocolumn]{revtex4}
\usepackage{amsmath}
\usepackage{bbm}
\usepackage{hyperref}
\usepackage{graphicx}
\usepackage{float}
\usepackage{color}
\usepackage{soul}
\usepackage{multirow}

\begin{document}
\title{Density functional plus dynamical mean-field theory of the metal-insulator transition in early transition-metal oxides}
\author{Hung T. Dang}
\affiliation{Institute for Theoretical Solid State Physics, JARA-FIT and JARA-HPC, RWTH Aachen University, 52056 Aachen, Germany}
\author{Xinyuan Ai}
\affiliation{Department of Physics, Columbia University, New York, New York 10027, USA}
\author{Andrew J. Millis}
\affiliation{Department of Physics, Columbia University, New York, New York 10027, USA}
\author{Chris A. Marianetti}
\affiliation{Department of Applied Physics and Applied Mathematics, Columbia University, New York, New York 10027, USA}

\begin{abstract}
The combination of density functional theory and single-site dynamical mean-field theory, using both Hartree and  full continuous-time quantum Monte Carlo impurity solvers, is used to study the metal-insulator phase diagram of perovskite transition-metal oxides of the form $AB$O$_3$ with a rare-earth ion $A$=Sr, La,  Y and transition metal $B$=Ti, V, Cr. The correlated subspace is constructed from atomiclike $d$ orbitals defined using maximally localized Wannier functions derived from the full $p$-$d$ manifold; for comparison, results obtained using a projector method are also given. Paramagnetic DFT+DMFT computations using  full charge self-consistency along with the standard ``fully localized limit'' (FLL) double counting are shown to incorrectly predict that LaTiO$_3$, YTiO$_3$,  LaVO$_3$ and SrMnO$_3$ are metals. A more general examination of the dependence of physical properties on the mean $p$-$d$ energy splitting, the occupancy of the correlated $d$ states, the double-counting correction, and the lattice structure demonstrates the importance of charge-transfer physics even in the early transition-metal oxides and elucidates the factors underlying the failure of the standard approximations.  If the double counting is chosen to produce a $p$-$d$ splitting consistent with experimental spectra, single-site dynamical mean-field theory provides a reasonable account of the materials properties. The relation of the results to those obtained from  ``$d$-only'' models in which the correlation problem is based on the frontier orbital $p$-$d$ antibonding bands is determined. It is found that if an effective interaction $U$ is properly chosen the $d$-only model provides a good account of the physics of the $d^1$ and $d^2$ materials.
\end{abstract}

\maketitle

\section{Introduction \label{Introduction}}

Understanding the ground state and excitations  of interacting electrons in solids is one of the grand challenges of modern condensed matter physics. The entanglement of coordinates in the fermion wave function imposed by the combination of Fermi statistics and the electron-electron interaction renders a solution of the all-electron many-body problem prohibitively difficult; indeed, theoretical arguments suggest that the general case of the many-electron problem is nondeterministic polynomial (NP) hard, meaning that it cannot be solved in polynomial time \cite{Troyer05}. While density functional theory (DFT) calculations \cite{Jones89} provide a useful and reasonably accurate treatment of many properties of many materials, in important cases such as  transition-metal oxides with partially filled $d$ shells DFT calculations often  fail \cite{Imada98} to provide even a qualitatively reasonable picture of the electronic properties of interest. ``Beyond-DFT'' electronic structure methods are needed.

In recent years the combination of density functional theory (DFT) and dynamical mean-field theory (DMFT) \cite{Georges04,Kotliar06,Held06} has emerged as a widely used beyond-DFT method. The approach has provided important qualitative insights into the  physics of important classes of materials including lanthanides and actinides \cite{Savrasov00,Marianetti08,Haule10}, transition metals \cite{Lichtenstein01}, transition-metal oxides \cite{Rozenberg95,Pavarini04,Biermann05,Park12} and many other compounds. One can formally view this approach as a dual-variable effective action theory where one constructs a functional of both the density and a local Green's function representing degrees of freedom in  a local subspace where correlations are most important \cite{Marianetti08}.  Two key issues remain imperfectly understood in this formally exact theory. The first issue is how to choose a local correlated subspace such that the best possible approximation can be developed when actually implementing the theory. This choice should be informed by the  approximations used in implementing the theory.  One commonly employed choice is to construct the correlated subspace  from frontier (near Fermi-surface) orbitals such as the $p$-$d$ antibonding bands of transition-metal oxides; examples may be found in Refs.~\cite{Rozenberg95,Pavarini04}. An alternative, and also widely used, choice is to define a correlated subspace in terms of atomiclike orbitals such as transition metal $d$ orbitals defined by applying a projector or Wannier construction to Kohn-Sham eigenfunctions in a wide energy range (see e.g.  Refs.~\cite{Vaugier12,Park12,Haule13}). We directly compare these two approaches in this study.

The second issue concerns the structure of the local potential that acts on the correlated substance.  While the theory is formally defined once the correlated subspace is chosen, in practical calculations  one must make approximations to the position dependent potential and the local time dependent potential (i.e., the self-energy) acting on the correlated subspace. These choices are the analogues of the choice of density functional in standard DFT. The local self-energy is obtained using the single-site dynamical mean field approximation \cite{Georges96}. Given the successes of the local density approximation (LDA)  and the generalized gradient approximation (GGA) \cite{Jones89,Perdew92}, it is natural to continue to use these approximations for the effective single-particle potential. The resulting formalism is termed the DFT+DMFT methodology \cite{Georges04,Kotliar06,Held06}. An obvious problem then arises, because the  LDA/GGA  exchange-correlation potentials already account for the local correlations to some degree.  Hence there is a ``double-counting'' problem, which needs to be corrected.  This  ``double-counting correction'' has been the subject of a considerable theoretical literature \cite{Anisimov91,Czyzyk94,Amadon08,Karolak10,Nekrasov12}  but remains ill-understood.  

In this paper, we study these issues via a detailed examination of the application of the DFT+DMFT methodology to the ``early'' transition-metal oxides. These materials crystallize in variants of the $AB$O$_3$ perovskite structure. The $B$ site contains an atom (Ti, V, Cr)  drawn from the left side of the first transition-metal row of the periodic table and the choice of $A$-site ion controls the filling of the $d$ level and aspects of the crystal structure. The early transition-metal oxides play a fundamental role in our understanding of the correlated electron problem, in particular exhibiting correlation-driven insulating states and the metal-insulator transitions (MIT) \cite{Imada98} that are not understandable in conventional density functional terms.  Elucidating the physics of these materials is a crucial step towards a more comprehensive solution of the many-electron problem, and understanding the factors controlling the  DFT+DMFT description of the materials is a crucial step in the validation of the method.  Our results demonstrate the importance of charge transfer physics even in the early transition-metal oxides and suggest that one issue with the DFT+DMFT program is that the underlying DFT provides an incorrect estimate of the charge transfer energetics, which then propagates into the many-body theory. We show that if this issue is corrected, then for electronically three dimensional materials the single-site dynamical mean field approximation provides a reasonably good approximation to the physics. Put differently, the uncertainty arising from our lack of knowledge of  the double-counting correction is apparently larger than the errors arising from the single site approximation to dynamical mean-field theory. 

The rest of this paper is organized as follows. In Sec.~\ref{sec:overview} we present a review of the DFT+DMFT method that emphasizes  the important physics issues. In Sec.~\ref{sec:model_method}, we present the theoretical model and  methodology used in this paper.  In Sec.~\ref{sec:hartree}, we present a simple but revealing Hartree-approximation solution to the DMFT impurity problem. This Hartree approximation provides a computationally  efficient method of  understanding the qualitative features of the phase diagram of various early transition-metal oxides, in addition to allowing a detailed comparison between correlated subspaces constructed from Wannier functions and from projectors. Section~\ref{sec:dmft} presents our DFT+DMFT results. In Sec.~\ref{sec:positions}, we discuss how to determine realistic values for the interaction and the double-counting. We consider the relationship of our results to those obtained by applying correlations to the frontier orbitals in Sec.~\ref{sec:donly}. Section~\ref{sec:conclusions} is a summary and conclusion.

\section{DFT+DMFT in Transition Metal Oxides\label{sec:overview}}

In this section we review two of the crucial technical and conceptual issues that arise in applying the DFT+DMFT method to transition-metal oxides, namely the definition of the correlated subspace  and the double-counting correction, in order to motivate the formalisms investigated  in this paper. 

\begin{figure}[t]
    \centering
    \includegraphics[width=\columnwidth]{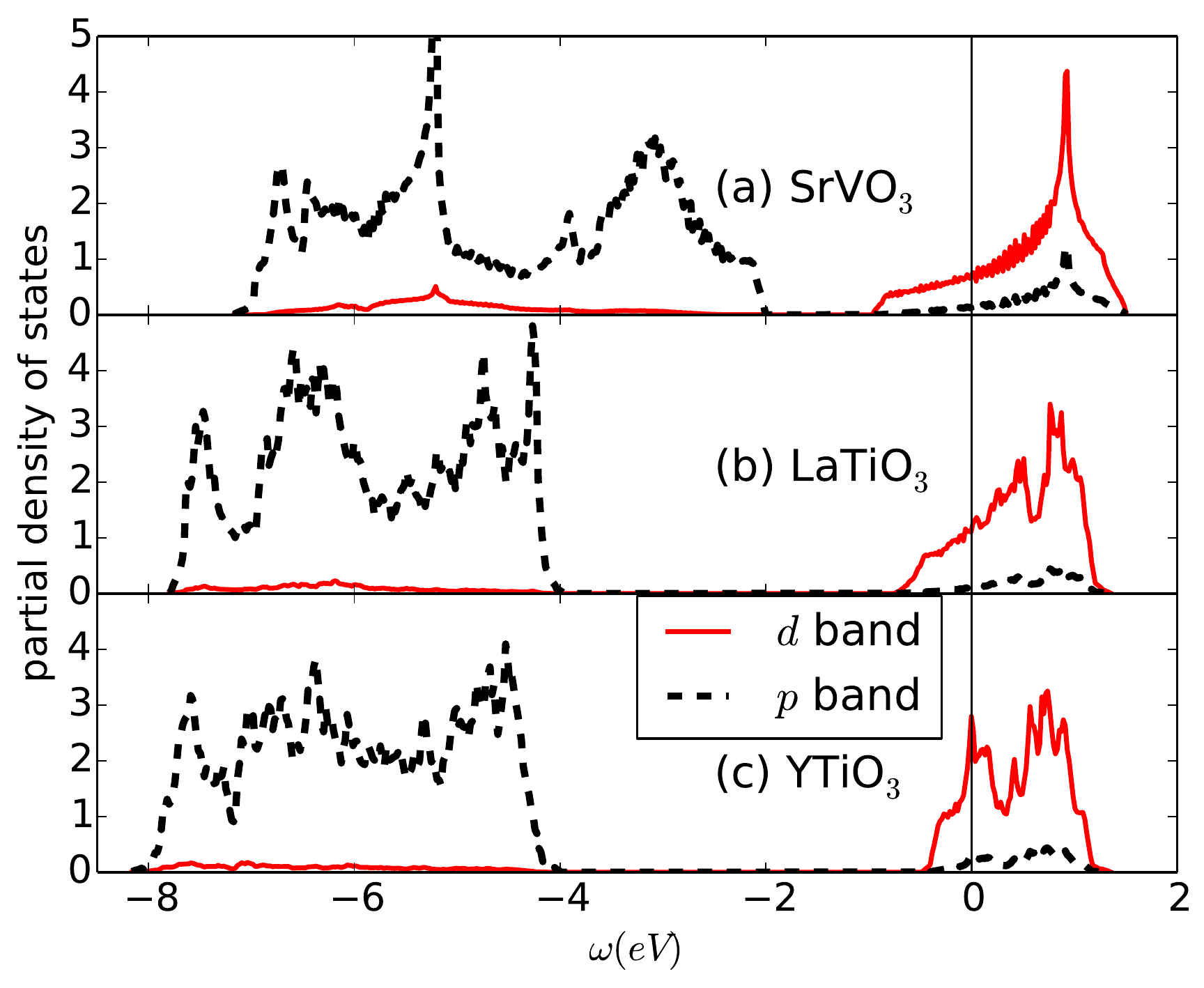}
    \caption{\label{fig:pdos_svo_lto_yto}(Color online) The density of states for SrVO$_3$, LaTiO$_3$ and YTiO$_3$ derived from DFT+MLWF tight binding Hamiltonian. The lattice structure for each material is from experimental data \cite{Rey90,Cwik03}. The vertical thin solid line marks the Fermi level. The solid curve (red online) is the transition metal $d$ band, the dashed curve (black online) is the oxygen $p$ band.}
\end{figure}

\subsection{Definition of the correlated subspace}
Applications of the DFT+DMFT method to transition-metal oxides are based on the idea,  accepted since the original work of Peierls and Mott \cite{Mott37,Mott1949},  that the appropriate correlated subspace consists of  the electrons in the partly-filled transition metal $d$ shell and that the important interactions to include in a beyond DFT calculation are the on-site, intra-$d$ interactions.  Different methods of constructing the correlated subspace have appeared in the literature.  In the early stages of theoretical development the correlated subspace was defined phenomenologically \cite{Hubbard63}, typically as a tight binding model of electrons hopping among sites of a lattice and coupled by an on-site repulsion $U$ and (if each site contains more than one orbital) a Hund's coupling $J$.  

Improvements in band structure calculations have made it possible to define the correlated orbitals  in a less phenomenological way. One widely adopted approach involves selecting the near Fermi-surface orbitals by fitting to a few-orbital  tight binding model using  downfolding \cite{Andersen95} or Wannier function \cite{PhysRevB.56.12847} techniques. This approach appears plausible for the early transition-metal oxides, where as seen in Fig.~\ref{fig:pdos_svo_lto_yto} the near Fermi-surface bands obtained from density functional calculations have a dominantly $d$-like character and are separated from the $p$ bands by an energy gap. DFT+DMFT studies based on this approximation have led to important insights; in particular Pavarini and co-workers have used this approach to demonstrate the crucial role played by GdFeO$_3$-type structural distortions in the metal-insulator transition of LaTiO$_3$ \cite{Pavarini04,Pavarini05} and have argued that the structural distortions are similarly important in LaVO$_3$ \cite{Raychaudhury07}. 

\begin{figure}
 \includegraphics[width=\columnwidth]{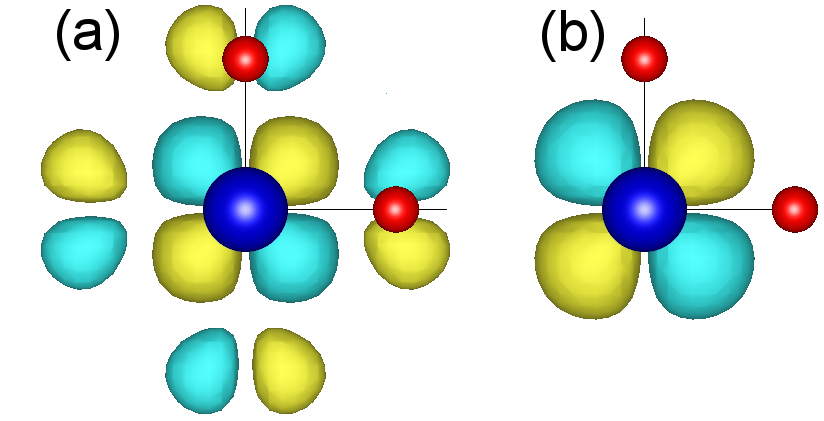}
\caption{\label{fig:mlwf_orbitals}(Color online) Energy isosurfaces of V-derived $d_{xy}$ orbitals of SrVO$_3$ (droplet shaped shaded regions, cyan and yellow online) along with positions of V and O ions (large circle, blue online, and small circle, red online, respectively). (a)  Frontier orbital used in the $d$-only model  defined by applying the maximally localized Wannier function construction to the near Fermi surface antibonding bands, (b)  atomiclike $d_{xy}$ orbital  used in the full DFT+DMFT procedure, defined by applying the maximally localized Wannier function method to the full $p$-$d$ band complex. The two plots are set at the same scale and the same isosurface value ($0.6$) and are generated using VESTA \cite{Momma11} with the input produced by Quantum Espresso \cite{QE-2009,QEPseudo} and Wannier90 \cite{Mostofi2008685}.}
\end{figure} 

Defining the $d$ orbitals via the near Fermi-surface bands however poses practical and conceptual difficulties. First, it seems desirable to have a theory that accounts in a unified way for the properties of all of the transition-metal oxides. However,  as the  $B$-site ion is varied across the $3d$ transition metal row from the ``early'' ions Ti and V to the ``late'' ions Ni and Cu, the admixture of $p$ states into the frontier bands increases and the  energy separation between the $p$ and the $d$ bands decreases, so an unambiguous identification of the frontier ($p$-$d$ antibonding) bands becomes problematic.  Second, even in the case of early transition-metal oxides, the real-space orbitals obtained from the near Fermi surface bands are rather delocalized in space, with orbitals centered on a transition metal ion having significant weight on the nearby oxygen ions and even some weight on the nearest-neighbor transition-metal ion (see Ref.~\cite{Andersen95} and  Fig.~\ref{fig:mlwf_orbitals} for examples). Whether the electron-electron interactions relevant to this somewhat delocalized object can be modeled with the simple  on-site $U$ and $J$ terms used in DFT+DMFT calculations is not clear.  Aichhorn and collaborators \cite{Aichhorn09} observed that projecting the physical interactions onto these frontier  orbitals let to an interaction matrix with symmetry properties  inconsistent with those found in the usual  theory of interactions in a $d$ shell.  A third issue relates to  the  ``charge transfer insulator'' physics introduced by  Zaanen, Sawatzky and Allen  in 1985. These authors observed \cite{Zaanen85} that if the energy $E_{CT}$ to transfer a charge from a ligand $p$ state to a transition metal $d$ state was less than the $d$ level charging energy $U$, then the physics is controlled by $E_{CT}$ and not $U$. While the precise ratio of $E_{CT}$ to $U$ that places the material in the charge transfer insulator regime is not known, we shall see that  descriptions of the physics of even the early transition-metal oxides require a $U\geq 4\text{eV}\gtrsim E_{CT}$, suggesting that charge transfer physics may be relevant even in these materials.

A more phenomenological reason why a focus on the frontier orbitals might be inadequate is seen in Fig.~\ref{fig:pdos_svo_lto_yto}. SrVO$_3$ is a moderately correlated metal, LaTiO$_3$ is a small-gap correlated insulator and YTiO$_3$ is a wider-gap correlated insulator. From Fig.~\ref{fig:pdos_svo_lto_yto} we see that the gap between oxygen $p$ bands and transition metal $d$ bands of SrVO$_3$ is about $1$eV. On the other hand, in LaTiO$_3$ and YTiO$_3$, the gaps are larger ($\gtrsim 3.5$eV), implying less $p$-$d$ mixing. While the difference in $p$-$d$ splitting appears in the frontier orbital model as a difference in bandwidths, the large differences in $p$-$d$ splitting between compounds may have additional effects, which cannot be studied in a frontier orbital-only model. 

For these reasons a different definition of the correlated orbitals, corresponding more closely to the intuitive idea of an atomiclike $d$ state, may be appropriate.  Such orbitals may be constructed by applying Wannier \cite{PhysRevB.56.12847} or  projector \cite{Haule10,Amadon08} methods to the set of states spanning the entire $p$-$d$ band complex. Provided that the manifold of states is defined over the full $p$-$d$ manifold, the correlated orbitals generated by either the projector or the Wannier procedure  are found to correspond reasonably closely to the intuitive picture of  atomic $d$ orbitals, having only small weight on the nearest neighbor oxygen ions (see, e.g. Fig.~\ref{fig:mlwf_orbitals}b). The two choices have been shown to lead to essentially  the same results in DFT+DMFT computations of La$_2$CuO$_4$ \cite{Wang12}. Further, for these orbitals the interaction matrix elements computed from constrained random phase calculations have, to a good approximation, the symmetry structure expected for the $d$ shell in free space  \cite{Aichhorn09}. Finally, the Zaanen-Sawatzky-Allen charge transfer physics can be included in the calculation on the same footing as the Mott-Hubbard physics driven by the local interactions. For most of this paper we adopt the atomiclike definition of $d$ orbitals, but in Sec.~\ref{sec:donly} we compare results obtained using frontier orbitals. 

\subsection{The double-counting correction}

Applying additional correlations to a predefined set of states creates a crucial complication: the extra correlations contribute  to a Hartree shift which will change the energies of the predefined states relative to other states in the material, and will therefore change the charge densities and other aspects of the physics.    In the transition metal oxide context, the extra correlations in particular change the energy of the $d$ level relative to that of the oxygen $p$ levels, shifting the charge-transfer energy $E_{CT}$ substantially and thus significantly affecting the Zaanen-Sawatzky-Allen metal-insulator transition physics while also changing the bandwidth and detailed band structure of the antibonding manifold.   Further,  a large change in $E_{CT}$ will lead to a large change in the occupancy of the  $d$ level, potentially  leading to issues with charge self-consistency by shifting the  charge distribution  away from the value favored by  the long-ranged Coulomb interaction. While some of the level shift may be physical (correcting errors in the underlying DFT), much of the Hartree shift associated with the physical on-site  correlations is included in the LDA/GGA estimates of the relative energies of the $p$ and $d$ states, and should not be counted twice. Therefore it is generally agreed \cite{Anisimov91,Czyzyk94,Karolak10,Nekrasov12}  that some forms of ``double-counting correction'' $\Delta$ should be introduced into the theory to  properly adjust the charge-transfer energy by compensating for some or all of the Hartree shift of the $d$ levels implied by the added interactions. Also,  a charge self-consistency process should be implemented to ensure that the long range part of the Coulomb energy is optimized. But because there is no clear theoretical procedure for deriving the double-counting correction, the literature has proceeded on a somewhat phenomenological basis, with different forms introduced based on symmetry and other arguments.  

Perhaps the most widely used form of the double-counting correction   is the fully localized limit (FLL) form \cite{Czyzyk94}
\begin{equation} \label{eq:fll_form}
E_{FLL}= U_{avg}\dfrac{N_d(N_d-1)}{2} - J_{avg}\sum_\sigma \dfrac{N_d^\sigma(N_d^\sigma-1)}{2},
\end{equation}
where $U_{avg}=\dfrac{1}{(2l+1)^2}\sum_{ij}U_{ij}$ and $U_{avg}-J_{avg}=\dfrac{1}{2l(2l+1)}\sum_{i\ne j}J_{ij}$. For the Slater-Kanamori interaction \cite{Kanamori1963} (see Eq.~\eqref{eq:onsite_SlaterKanamori}), $U_{avg}$ and $J_{avg}$ are:
\begin{equation}\label{eq:UJavg}
\begin{aligned}
 U_{avg} = U-\dfrac{8J}{5}, \hspace{12mm}
 J_{avg} = \dfrac{7J}{5}.
\end{aligned}
\end{equation}
We note that in VASP DFT+$U$ or the Wien2k/TRIQS code, the interaction is written in the form of the spherical harmonic functions \cite{0953-8984-9-4-002} (see Table~\ref{code_summary}), which is not identical to,  but can be well approximated by, the Slater-Kanamori form provided that the interaction parameters are such that both forms of interactions yield the same $U_{avg}$ and $J_{avg}$. Therefore, when using the VASP DFT+$U$ and Wien2K/TRIQS codes, we present our results in terms of the $U$ and $J$ implied by the  $U_{avg}$ and $J_{avg}$ via   Eq.~\eqref{eq:UJavg}.

The addition of $E_{FLL}$ to the functional yields a term $\Delta$ in the effective potential which shifts  the correlated subspace relative to the other electronic states:
\begin{equation}
\begin{split}
\Delta_{FLL}^{\sigma} &= U_{avg}\left(N_d-\frac{1}{2}\right) - J_{avg}\left(N_d^\sigma-\frac{1}{2}\right) \\
&= \left(U-\frac{8J}{5}\right)\left(N_d-\frac{1}{2}\right) - \frac{7J}{5}\left(N_d^\sigma-\frac{1}{2}\right)
\label{DeltaFLL}
\end{split}
\end{equation}

We  found \cite{Wang12} that for  La$_2$CuO$_4$, the FLL double counting in combination with the  fully charge self-consistent DFT+DMFT procedure yields metallic behavior, while the material is insulating in experiment. In subsequent work \cite{Park13}, Park and two of us  found that DFT+DMFT calculations in conjunction with the FLL double-counting wrongly predicts that none of the rare earth nickelate family $R$NiO$_3$ have a  charge disproportionated ground state, whereas in experiment \cite{Alonso99} all of the materials except LaNiO$_3$ disproportionate. The essential difficulty was found to be that the FLL double counting places the $d$ levels too close to the $p$ levels. 

One possible resolution of this problem is to alter the double-counting correction. Park and two of the present authors \cite{Park13} proposed a modified FLL formula in which $U$ is replaced by a smaller value $U'<U$. This form was motivated by studies of the total energy within DFT+DMFT and can be used straightforwardly  to perform fully charge self-consistent calculations in existing codes. The effect of using a $U^\prime<U$ is to increase the $p$-$d$ splitting and thus slightly decrease the number of electrons in the correlated shell; for the LaNiO$_3$ system, this approach was found to give much better agreement with multiple experiments \cite{Park13}.  An alternative approach  is given in Ref.~\cite{Haule13}, which proposed that Eq.~\ref{DeltaFLL} be replaced by a constant level shift determined by replacing ${N_d}$ in that equation with the formal valence $N_d^0$. This approach  has the practical effect of  reducing the magnitude of the  double-counting correction relative to the FLL value,  thereby increasing the $p$-$d$ splitting.  The ansatz of Ref.~\cite{Haule13} however implies that the double counting contribution to the Hamiltonian   should not be viewed as an interaction energy, but  that Eq.~\eqref{eq:fll_form} should be replaced by a linear function of ${N_d}$. 

Other approaches have also been discussed \cite{0953-8984-9-4-002,Amadon08,Karolak10,Nekrasov12}, but the theoretical issue is not settled. In previous work \cite{Wang12} we therefore proposed  to sidestep entirely the question of what form of double-counting correction should be used. We demonstrated that for cuprates and nickelates different choices of double-counting correction correspond in the end to different values of the charge-transfer energy or, equivalently, to different values of  the number $N_d$ of electrons in the correlated shell. To understand the physics of the metal-insulator transitions, we computed the metal-insulator phase diagram for theoretical models of La$_2$CuO$_4$ and LaNiO$_3$ as a function of $U$ and $\varepsilon_d-\varepsilon_p$ and presented the results  in the plane of $U$ and ${N_d}$, revealing that for $N_d$ sufficiently close to the nominal formal valence value $N_d^0$ (for example $N_d^0=9$ for cuprates) the model is insulating while when $N_d$ exceeds a critical value, an insulator to metal transition ensues. An interesting aspect of this representation of the data is that for large $U$ the phase boundary generically becomes nearly vertical, indicating that for sufficiently large $N_d$ (i.e., for sufficiently small charge-transfer energy) an insulating state cannot be realized even for large values of $U$.  

In this paper, we examine the extent to which these issues are relevant to a wider range of transition-metal oxides, in particular the ``early'' transition-metal oxides such as the La and Sr-based  titanates, vanadates, and chromates.

\section{Methods \label{sec:model_method}}

\subsection{Overview}
In this paper we shall mainly be interested in transition-metal oxides that crystallize in variants of the $AB$O$_3$ perovskite structure. We study materials in which the $A$-site ion is Sr, La and (in one case) Y and the $B$ site ion is one of Ti, V, Cr and Mn. The Sr series of materials are cubic perovskites; the La/Y series crystallize in GdFeO$_3$-distorted variants of the cubic perovskite structure characterized by a four-sublattice pattern of tilts and rotations.

\begin{table}
\begin{ruledtabular}
\begin{tabular}{ccccc}
  \multirow{2}{*}{Code} & Correlated & Impurity & \multirow{2}{*}{CSC} & \multirow{2}{*}{Interactions} \\
                        & Subspace   & Solver   &     &  \\
  \hline                       
  Quantum Espresso      & MLWF       & CT-QMC, Hartree & No & SK \\
  VASP         & Projector & Hartree & Yes & SH \\
  Wien2K/TRIQS & Projector & CT-QMC  & Both  & SH \\
\end{tabular}
\end{ruledtabular}
\caption{ \label{code_summary} A summary of the DFT  codes used (Quantum Espresso code \cite{QE-2009,QEPseudo}, VASP \cite{Kresse93,Kresse96a,Kresse96b,Kresse99}, and Wien2k \cite{Wien2k}), the methods (maximally localized Wannier function (MLWF)  \cite{PhysRevB.56.12847,PhysRevB.65.035109} as implemented in Wannier90 \cite{Mostofi2008685}, or projector \cite{Haule10,Amadon08}) employed to construct the correlated subspace, the impurity solver, whether or not full charge self-consistency (CSC) is implemented, and whether the Slater-Kanamori (SK, Eq.~\eqref{eq:onsite_SlaterKanamori}) or spherical harmonic (SH, Ref.~\cite{0953-8984-9-4-002}) forms of the interaction are used. We note that the projectors defined in VASP and Wien2K/TRIQS have different implementations.}
\end{table}

In this study we have used  different DFT codes,  methods of constructing the correlated subspace, impurity solvers and forms of the interaction in order to obtain an understanding of the effect of these details on the physics.  A summary showing which  methodological options were used with each code is given in Table~\ref{code_summary}.

We will demonstrate that for the V and Ti-based compounds, it is not necessary to treat the entire $d$-manifold. Truncation to the $t_{2g}$ subspace provides an accurate representation of the physics: the $e_g$ levels (which are nearly empty) may be omitted entirely. However, for LaCrO$_3$ and SrMnO$_3$ where the standard valence counting indicates that the $t_{2g}$ shell is half filled, we find that inclusion of the $e_g$ levels is important, essentially because the insulating gap is determined by the energy difference between the $t_{2g}$ and $e_g$ levels.

The interactions in the correlated subspace are normally taken as the standard Slater-Kanamori form \cite{Kanamori1963}
\begin{equation}\label{eq:onsite_SlaterKanamori}
\begin{split}
H_{onsite} & = U\sum_{\alpha}n_{\alpha\uparrow}n_{\alpha\downarrow}  + (U-2J)\sum_{\alpha\neq\beta} n_{\alpha\uparrow}n_{\beta\downarrow} + \\
& + (U-3J)\sum_{\alpha > \beta,\sigma}n_{\alpha\sigma}n_{\beta\sigma} + \\
& + J\sum_{\alpha\neq\beta} ( c^\dagger_{\alpha\uparrow}c^\dagger_{\beta\downarrow}c_{\alpha\downarrow}c_{\beta\uparrow}
+ c^\dagger_{\alpha\uparrow}c^\dagger_{\alpha\downarrow}c_{\beta\downarrow}c_{\beta\uparrow}
).
\end{split}
\end{equation} 
where $\alpha,\beta$ label orbitals in the transition-metal $d$ manifold on a given site. We fix $J$, which is only very weakly screened by solid state effects, to be $J=0.65$eV unless stated otherwise, but consider a range of $U$.

\subsection{Solution of correlation problem\label{sec:model:corr_solution}}

We obtain the local self-energy using the single-site dynamical mean field approximation \cite{Georges96}, which requires the solution of an auxiliary quantum impurity  model. We obtain  numerically accurate  solutions using  quantum Monte Carlo methods \cite{Werner06,Gull11} and also simple and qualitatively useful approximations using the Hartree method, in which the quartic terms of the Hamiltonian are approximated by density mean fields $ \langle n_i\rangle $ determined self consistently such that
$
n_i n_j \approx n_i\langle n_j\rangle + \langle n_i\rangle n_j - \langle n_i\rangle \langle n_j\rangle.
$
It should be noted that the DFT+DMFT formalism reduces to DFT+$U$ when solving the DMFT impurity problem within the Hartree approximation.  We also note that while the spin polarization is allowed in DFT+$U$ or DFT+Hartree, all calculations with DFT+DMFT in this work are restricted to the paramagnetic state.

\begin{figure}[t]
    \centering
    \includegraphics[width=\columnwidth]{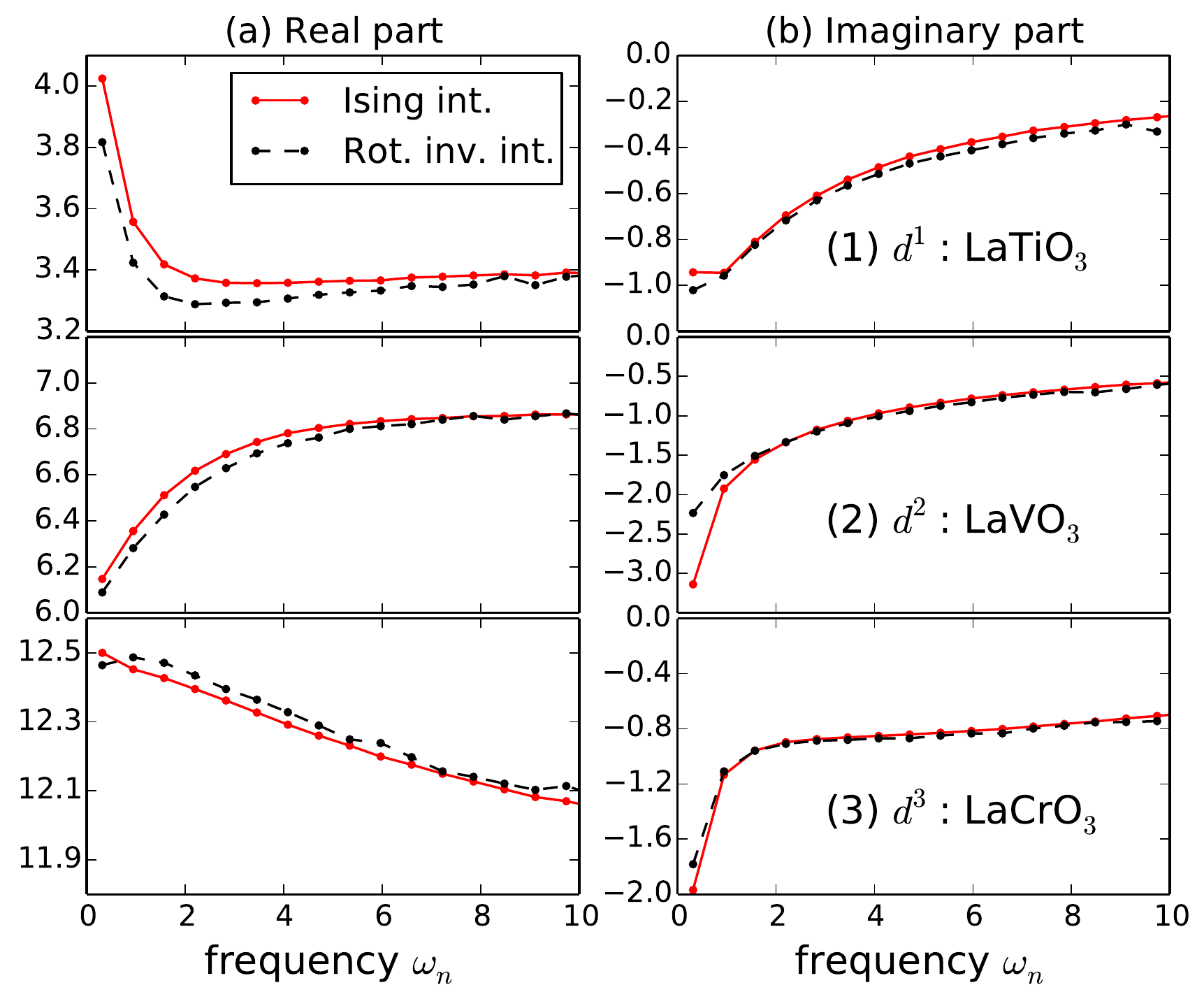}
    \caption{\label{fig:ising_rotinv}(Color online) Imaginary part of the Matsubara self energies obtained with $U=5$eV and $J=0.65$eV using  Ising interaction (pair hopping and exchange terms excluded) and rotationally invariant interaction and plotted against Matsubara frequency.  The $p$-$d$ energy splitting is the same for the Ising and rotationally invariant calculations but  is adjusted for each material so that the compound is near the metal-insulator phase boundary (see Fig.~\ref{fig:dmft_la_series}).}
\end{figure}

In order to perform the extensive calculations needed for our phase diagram surveys we typically neglect the exchange and pair-hopping terms of Eq.~\eqref{eq:onsite_SlaterKanamori} (Ising approximation) in our QMC calculations, to be able to use the ``segment'' algorithm (see Ref.~\cite{Gull11} for a definition), which is 4 to 5 times faster. To test the quality of the interaction we present in Fig.~\ref{fig:ising_rotinv}  a comparison of the  self-energy obtained using the  rotationally invariant and Ising interactions for $d^1$, $d^2$ and $d^3$ systems with $U=5$eV and $N_d$ chosen so that the materials are near the metal-insulator phase boundary. The imaginary part of the self-energy, which is a reasonable representation of the correlation strength, is similar in the two cases, except at the lowest frequency. The differences in self-energy are found to be sufficiently small so that the metal-insulator phase diagram is well approximated by the Ising interaction calculations.

To specify whether the system is metallic or insulating, we use maximum entropy techniques to continue the self-energy, then use the continued self-energy to compute the lattice Green's function.  We define solutions as insulating if the imaginary part of the local Green's function vanishes at the chemical potential. To locate the metal-insulator transition phase boundary, we determine the gap magnitude from a linear extrapolation of the density of states and define the metal-insulator transition as the point at which the gap is closed.

\subsection{The double-counting correction, full charge self-consistency and the $d$ level occupancy}
The double-counting correction in effect defines a shift $\Delta$ of the correlated subspace that   acts to compensate for some or all of  the Hartree shift due to the interactions within the subspace.  Different forms of the double-counting correction have been given in the literature \cite{Anisimov91,Czyzyk94,Amadon08,Karolak10,Nekrasov12,Haule13} but the correct form is not known. Determining the correct form of the double-counting correction (or, alternatively, the correct mean $p$-$d$ energy splitting) is a crucial open issue in the DFT+DMFT methodology. 

In the fully charge self-consistent DFT+$U$ and DFT+DMFT calculations we use the fully-localized-limit (FLL) double counting formula \cite{Czyzyk94} [see Eq.~\ref{DeltaFLL}] unless otherwise stated. In our other calculations (which do not include the charge self-consistency step), we follow Ref.~\cite{Wang12} and consider a range of double-counting corrections, which we parametrize by $N_d$, the expectation value of the operator giving the $d$ level occupancy. The parametrization is possible because if the correlation problem is defined in terms of the $p$-$d$ manifold,  $N_d$ is a monotonic function of the $d$ level energy. The parametrization is useful because (as was demonstrated for late transition-metal oxides in Ref.~\cite{Wang12} and will be seen in detail below) many of the specifics  of the materials properties affect the metal-insulator line only via their effect on the value of $N_d$, so the resulting phase diagrams are relatively simple when expressed in terms of $N_d$. Of course the precise values found for  $N_d$ depend on the precise definition of the $d$ orbital which in turn depends on the scheme (Wannier versus projector) and the energy window chosen. However the trends are robust and different situations can be meaningfully compared if consistent definitions of $d$ orbital are adopted. Further details are given in Ref.~\cite{Wang12}. 

A related important issue concerns the effect of  full charge self-consistency in the DFT+DMFT formalism. In Ref.~\cite{Dang13} we showed that the only important effect of the full charge self-consistency is a change in the $N_d$ values: a one-shot calculation tuned to have the same $N_d$ produces spectra that are indistinguishable from those obtained in the fully charge self-consistent formalism. We present here additional calculations further confirming this observation. Therefore, in most of this paper we simply present non-charge self-consistent results as a function of $N_d$.

\subsection{GdFeO$_3$ distortion}

In reality, only a few perovskites (most notably SrVO$_3$ and SrMnO$_3$) form in the cubic structure. In most transition-metal oxides of chemical form $AB$O$_3$, the small radius of the rare earth $A$ causes a significant distortion (the GdFeO$_3$ rotation)  of the perovskite structure. Pavarini and collaborators \cite{Pavarini04} argued on the basis of studies of a frontier orbital ($d$-only) model that the distortion was  important for  the metal-insulator transition. The importance of the crystal structure was also noted by Craco \textit{et al.} \cite{Craco04}. We further investigate this issue using our approach. 

The materials we study form in the $Pnma$  or in Glazer's notation \cite{Glazer72} $a^-b^+c^-$  structure. These structures may be obtained from the ideal perovskite structures by rotating the transition-metal-oxygen octahedra by certain tilt angles. The tilt angles are zero for SrVO$_3$ and increase as one moves to LaTiO$_3$, LaVO$_3$ and finally to YTiO$_3$. The experimental structural parameters \cite{Rey90,Cwik03,Bordet93} are used in all calculations.

\begin{figure*}[t]
    \centering
    \includegraphics[width=0.45\textwidth]{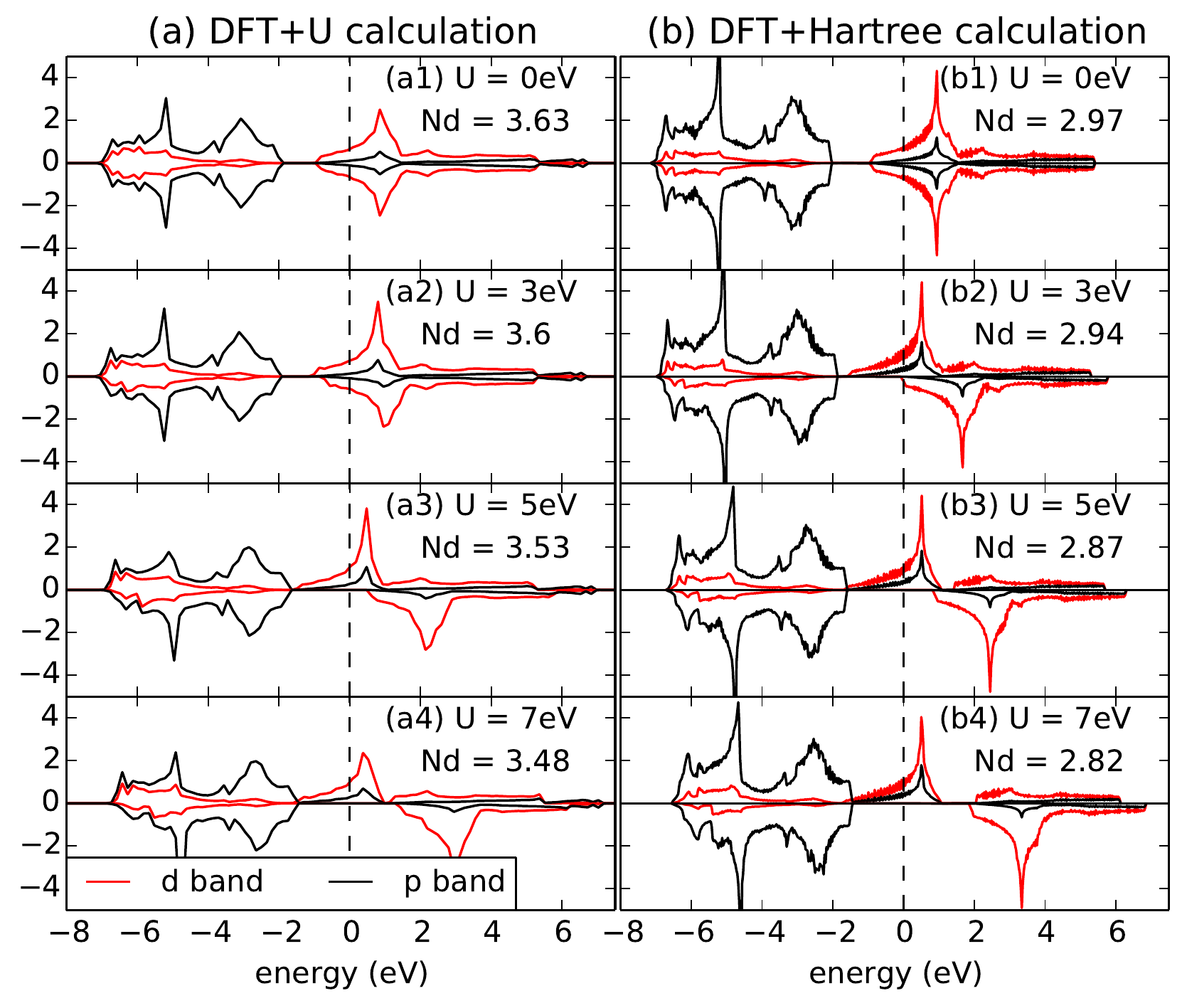}
    \includegraphics[width=0.45\textwidth]{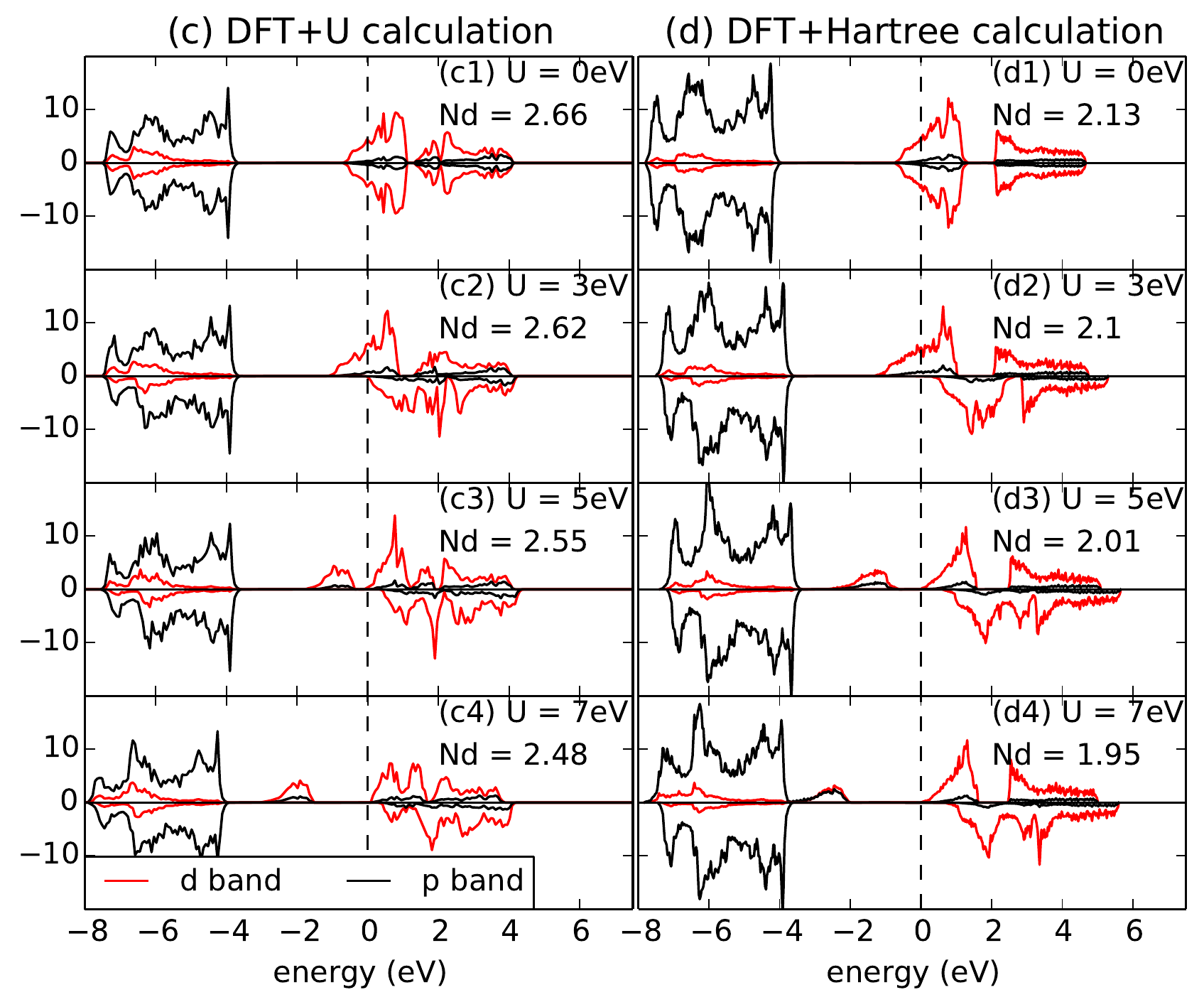}
    \caption{\label{fig:pdos_dftnd}(Color online) The density of states for SrVO$_3$ (SVO) and LaTiO$_3$  (LTO)  obtained from DFT+$U$ (VASP implementation) with FLL double counting [(a) for SVO and (c) for LTO] and DFT+Hartree (Quantum Espresso/MLWF) [(b) for SVO and (d) for LTO] for $U=0,3,5$ and $7$eV. The DFT calculations employ the experimental structures.  The Hund's coupling is $J=0.65$eV. The light (red online) curves are the transition metal $d$ bands, the black curves are the oxygen $p$ bands. The $N_d$ values shown in the DFT+$U$ columns are calculated from the VASP projector, while the ones in the DFT+Hartree columns are from MLWF. The dashed lines mark the Fermi level, which is set at the lower edge of the majority spin upper band. In the DFT+Hartree calculations, the double-counting correction is manually set so that the decrease in $N_d$ is the same as in DFT+$U$ calculations as $U$ increases.}
\end{figure*}

While the hybridization function is diagonal in orbital indices for the cubic structures,  it will have off-diagonal terms in the GdFeO$_3$ structures. Because off-diagonal terms in the hybridization function lead to a severe sign problem in the CT-QMC calculations it is advantageous to define a basis in which the off-diagonal terms are minimized. We therefore  employ on each site a new (rotated) basis of  $t_{2g}$ orbitals chosen to minimize off-diagonal terms \cite{Dang13b}. In our procedure, at each DMFT iteration, the lattice Green's function is rotated to the new basis in order to obtain a diagonal hybridization function which serves as the input of the impurity solver. The output diagonal impurity self-energy is then transformed back to the original basis in preparation for the next DMFT iteration. The results from the DMFT calculation are post-processed in the same ways as for the cubic structure to construct the MIT phase diagrams.

\section{DFT+Hartree calculations\label{sec:hartree}}

In this section we solve the DMFT impurity problem within the Hartree approximation as described in Sec.~\ref{sec:model:corr_solution} (based on Quantum Espresso/MLWF) and compare the results to those obtained with the widely used DFT+$U$ approximation as implemented in VASP \cite{Kresse93,Kresse96a,Kresse96b,Kresse99}. The VASP DFT+$U$ calculations are based on a definition of the $d$ orbitals from a projection onto $d$-symmetry states defined within a sphere centered on the transition metal sites and include a full charge self consistency calculation with  the FLL double counting [Eq.~\ref{DeltaFLL}]. The spin-independent PBE \cite{Perdew96} exchange-correlation functional is employed. Hereafter we will refer to these two approaches as DFT+Hartree and DFT+$U$, but it should be understood that the only difference is that the former uses a correlated subspace defined via Wannier functions and does not perform full charge self-consistency  while the latter has a correlated subspace defined via a projector and does include full charge self-consistency. 

\begin{figure*}[t]
    \centering
    \includegraphics[width=0.45\textwidth]{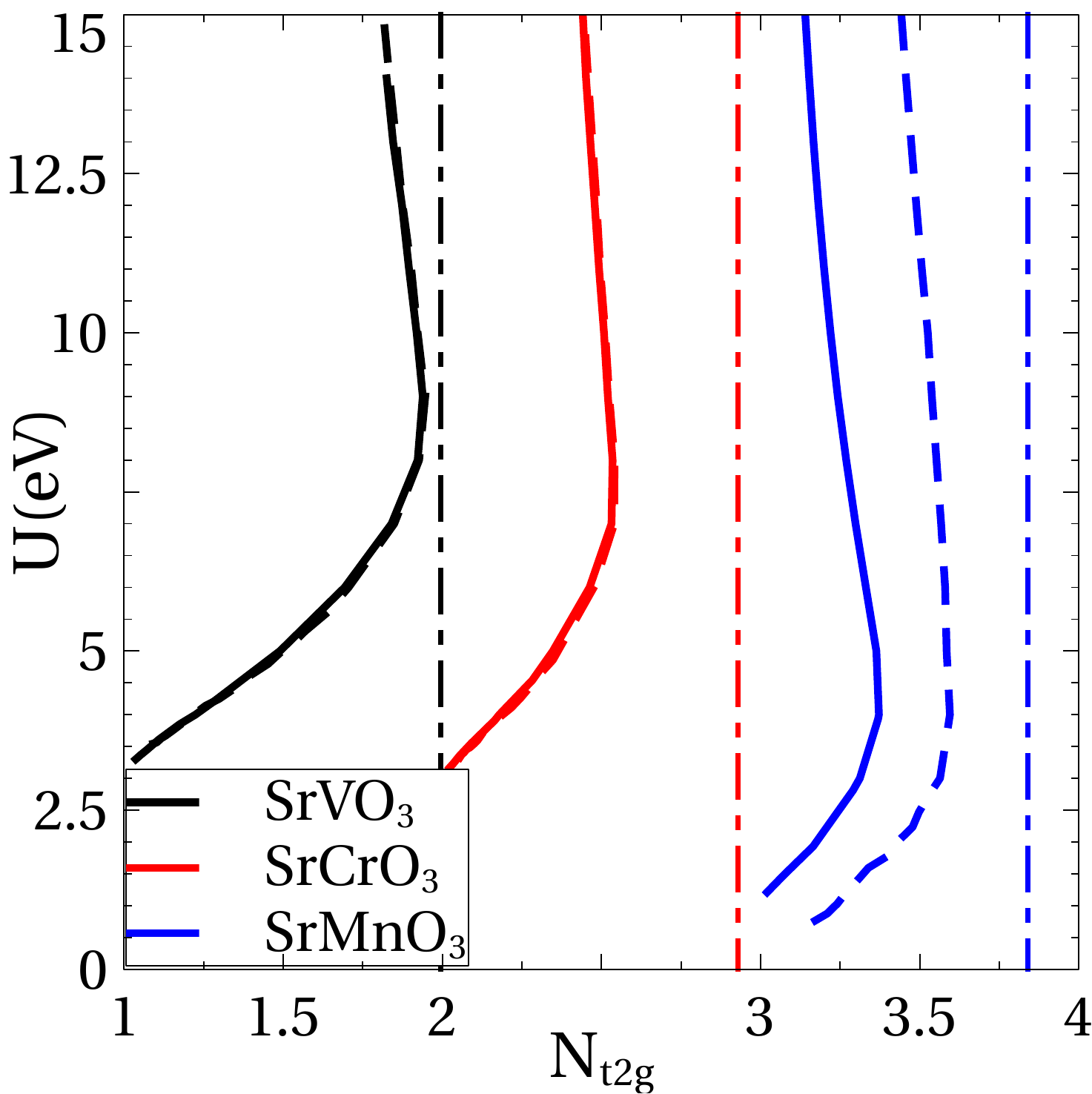}
    \includegraphics[width=0.45\textwidth]{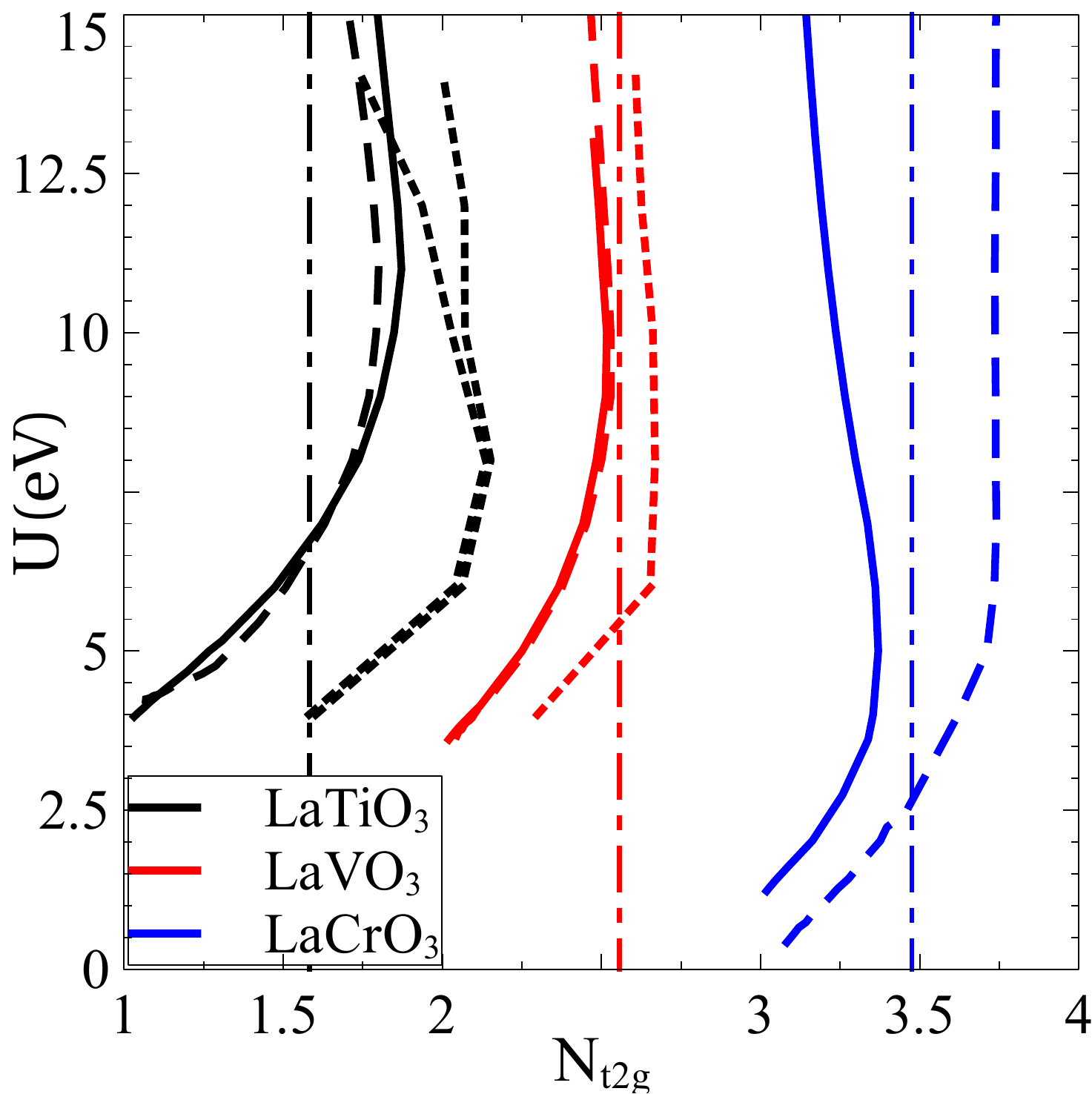}
    \caption{\label{fig:s_l_comp}(Color online) Metal-insulator phase boundary computed using the DFT+Hartree approximation for the Sr and La-based perovskites in the cubic and GdFeO$_3$-distorted structures and displayed in the plane  of $t_{2g}$ occupancy and interaction $U$, with the Hund's coupling $J=0.65$eV. The solid lines show the phase boundary computed using the full 5 $d$ orbital model while the dashed lines show the results obtained by restricting to the $t_{2g}$-only manifold. The dotted curves are the phase boundaries of LaTiO$_3$ and LaVO$_3$ using their real structures (LaTiO$_3$ has two dotted curves with an area of phase separation in between). The insulating (metallic) regime is to the left (right) of the phase boundary. The vertical lines mark the values of $t_{2g}$ occupancy obtained from density functional band calculations.}
\end{figure*}

The DFT+Hartree and DFT+$U$ calculations are static mean field approximations, and as such overemphasize the tendency to long ranged order and provide poor approximations to spectra. However, the methods are computationally inexpensive and provide important insights. In our calculations we do not allow the possibility of breaking of translational symmetry; therefore, insulating behavior requires ferromagnetic and ferro-orbital order.  Allowing for antiferromagnetic and/or antiferro-orbital states which break translational symmetry would change the locations of the phase boundaries, but would not affect the qualitative conclusions we wish to draw here, concerning the relation of Wannier and projector results, the effect of charge self-consistency, and the relevance of the $e_g$ manifold of states.

Figure~\ref{fig:pdos_dftnd} compares the VASP DFT+$U$ fully charge self consistent calculations (with FLL double counting) to those obtained from  DFT+Hartree approximation calculations in which the $d$ level energy is adjusted to produce spectra in agreement with the VASP DFT+$U$ results (in particular relative energies of the majority $p$ and majority $d$ derived bands). Even after this adjustment, small differences remain between the two calculational methods. We discuss these in detail below but emphasize that the small differences do not change any of the qualitative physics. 

The fully charge self consistent DFT+$U$ calculations in Fig.~\ref{fig:pdos_dftnd}  show that, as the interaction strength is increased, the electronic structure rearranges itself so as to keep  the $d$ occupancy and the $p$-$d$ band splitting (defined, for example, as the energy separation from the top of the lower, oxygen-dominated bands to the bottom of the majority spin upper band) relatively unchanged. We see that as $U$ is increased the  energy of the ``upper Hubbard band'' (minority spin unoccupied states) increases. This increase implies a decrease in virtual charge fluctuations into the minority spin $d$ states. However the decrease in $N_d$ implied by this decrease in virtual charge fluctuations is to a large extend compensated by a small upward shift of the O states (compare e.g. the position of the O states relative to the Fermi level in panels a1-a4 of Fig.~\ref{fig:pdos_dftnd}) which acts to increase the occupancy of the majority-spin orbitals, with the result that $N_d$ is hardly changed. This evolution of electronic structure with $U$ reveals an essential role of $p$-$d$ covalency and charge self-consistency in compensating for the effects of the Hubbard $U$. This physics  is not contained in the frontier orbital Hubbard model.

There are some  differences of detail  between DFT+$U$ and DFT+Hartree calculations. First, the projected DOS resulting from the  DFT+$U$ calculation is slightly smaller than that of the DFT+Hartree calculation, because  some portion of the charge resides in interstitial regions and is not captured  by the projector method used in the VASP implementation of DFT+$U$. Second, as can easily be seen by comparing the $d$ DOS  in the $p$-dominated lower energy part of the spectrum shown in Fig.~\ref{fig:pdos_dftnd} (the same effect is present but more difficult to discern in the $d$-dominated part of the DOS),  the $d$ occupancy resulting from the VASP DFT+$U$ calculation  is larger than that resulting from the MLWF-based DFT+Hartree procedure.  For SrVO$_3$ the total $N_d$  per transition metal atom (summed over all 5 $d$ orbitals) obtained in the VASP DFT+$U$ projector scheme is  about $0.65$ greater than that obtained in the DFT+Hartree Wannier scheme; for LaTiO$_3$ the difference is about  $0.53$. Roughly half of this difference arises from the fact that the projector has nonzero weight below $-8$eV. The more relevant contribution to the  difference arises because  the projector method produces a slightly larger $p$-$d$ covalency than the Wannier method. A consequence is that because the dimensionless parameter giving the effective correlation strength  of the quantum impurity model is more or less the ratio of the interaction $U$ to a measure of the covalence, the projector-based DFT+$U$ results are in effect less correlated than the DFT+Hartree results, explaining  the difference in gap sizes and spin polarizations between the two methods. Third, in DFT+$U$ calculations for LaTiO$_3$ [Fig.~\ref{fig:pdos_dftnd}c], the Ti $d$ bands mix with La $f$ bands, resulting in small portions of Ti $d$ DOS at the positions of La $f$ bands slightly above the Fermi level [see, for example, the DOS in the energy range between $1.5$ and $2$eV in panel 1 of Fig.~\ref{fig:pdos_dftnd}c]. This mixing is not captured in MLWF method used in DFT+Hartree [compare with panel 1 of Fig.~\ref{fig:pdos_dftnd}d]. These differences do not affect the qualitative trends and make only small changes to quantitative values but are important to bear in mind when comparing projector and Wannier-based results.

\begin{figure}[t]
    \centering
    \includegraphics[width=\columnwidth]{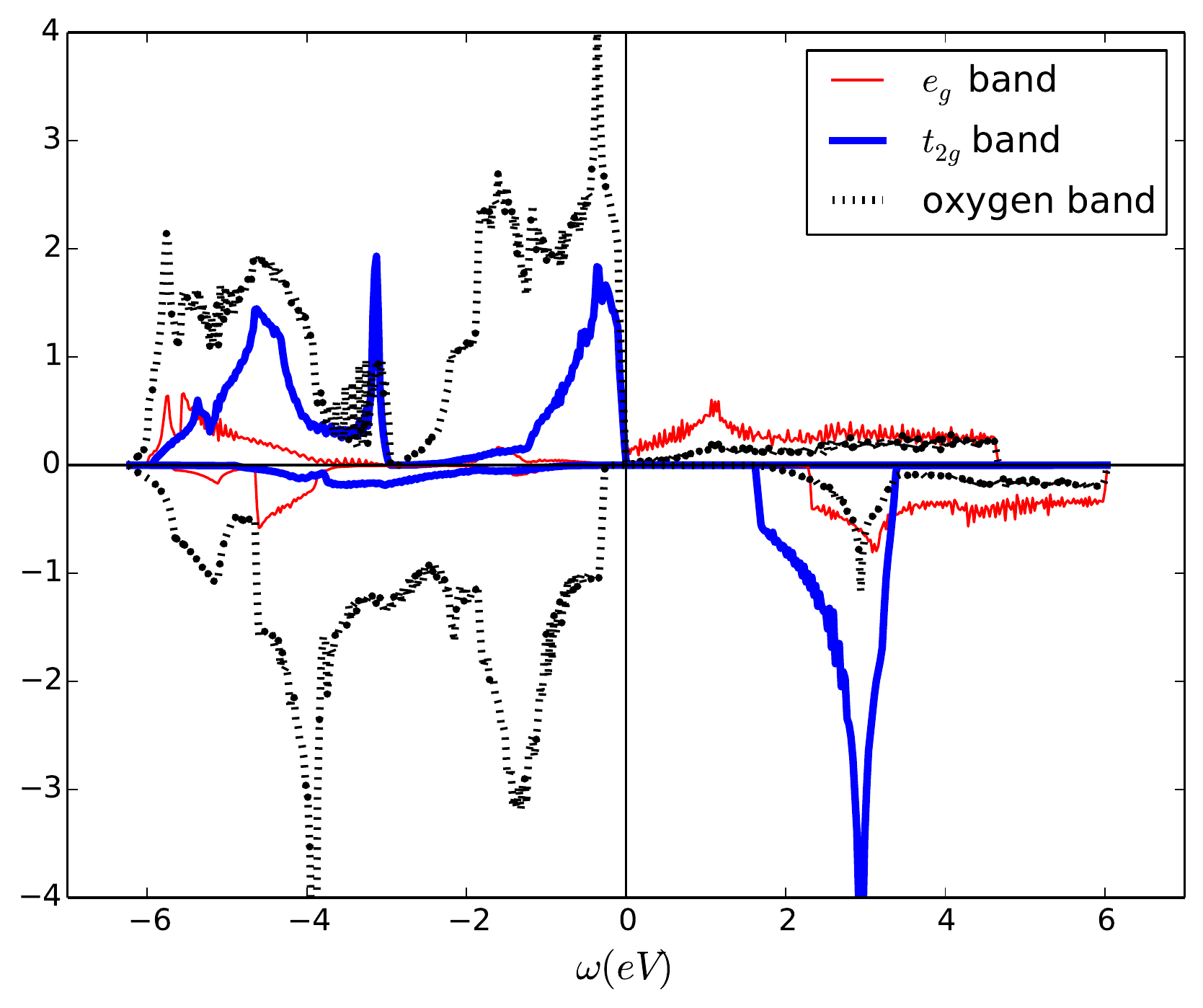}
    \caption{\label{fig:dos_smo_5bands}(Color online) Density of states for SrMnO$_3$ at $U=4$eV and $J=0.65$eV generated using DFT+Hartree assuming unbroken translational symmetry and  using five $d$ and nine oxygen $p$ orbitals. The double-counting correction is adjusted so that $N_d=3.3$ and the system is at the metal-insulator phase boundary. The solution is ferromagnetic with no orbital ordering. The positive (negative) DOS is for the majority (minority) spin. The vertical line marks the Fermi level.}
\end{figure}

Figure~\ref{fig:s_l_comp} shows the metal-insulator phase diagram computed in the DFT+Hartree approximation for the Sr and La series of materials. First we consider the simple cubic perovskite lattice structure; this is physical for the Sr series but not for the La series. As we do not allow translational symmetry breaking, obtaining insulating states with  DFT+Hartree calculation requires ferromagnetic (translationally invariant breaking of spin symmetry) and ferro-orbital (translationally invariant breaking of rotation symmetry about a transition metal site) order. Orbital order corresponds to splitting the energies of the $t_{2g}$ levels, which are  degenerate in the non-orbitally ordered state.  We find two kinds of splitting pattern: ``1 down, 2 up'', in which one orbital has lower energy than the other two and correspondingly higher occupancy,  and ``2 down, 1 up'' in which two orbitals have approximately the same energy, which is lower than that of the third, so that they have higher occupancy. We find that the orbital order depends on the  formal  valence: $d^1$ has ``1 down, 2 up'', $d^2$ has ``2 down, 1 up'' and $d^3$ has no orbital order.  We see that for both the Sr and hypothetical cubic La series of materials, in the nominally $d^1$ and $d^2$ compounds the three $t_{2g}$ orbital (the dashed curves) and five orbital (the solid curves) calculations yield essentially indistinguishable results, whereas in the nominally $d^3$ materials, inclusion of the $e_g$ manifold changes the physics significantly, drastically decreasing the parameter regime over which insulating behavior is found.  

To explicate the reason for the difference we show in Fig.~\ref{fig:dos_smo_5bands} the density of states computed in the DFT+Hartree calculation for SrMnO$_3$, with parameters tuned so that the system is in the metallic state but on the boundary of the insulating phase. It is evident that the gap is between the $t_{2g}$ and $e_g$ manifolds; inclusion of the $e_g$ states is thus essential to describe the physics. By contrast, in the nominally $d^{1,2}$ materials the excitation gap is to  unoccupied $t_{2g}$ states; $e_g$ states do not play an important role in the metal-insulator transition. Further, we observe that in contrast to the $d^1$ and $d^2$ systems, where the insulating gap is closely related to the $p$-$d$ splitting which is directly connected to $N_d$, in the $d^3$ systems the insulating gap arises from the $e_g$-$t_{2g}$ splitting which is affected only indirectly by $N_d$.

\begin{figure}[t]
    \centering
    \includegraphics[width=\columnwidth]{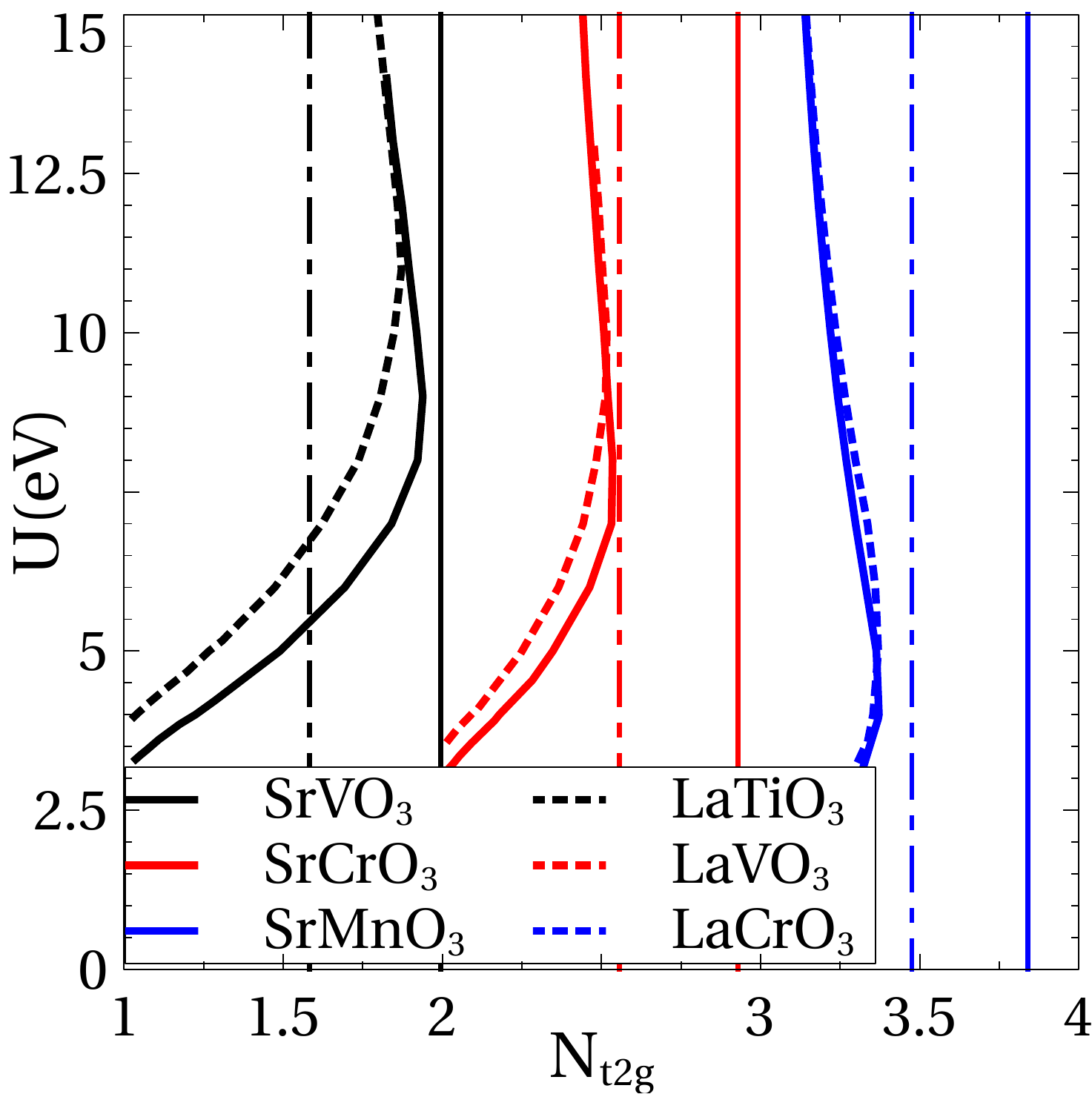}
    \caption{\label{fig:hf_phasediagrams}(Color online) Comparison of DFT+Hartree phase diagrams for Sr-based and La based perovskites. All calculations are performed for cubic structures. The full 5-$d$ orbital model is used. Hund's coupling is $J=0.65$eV.}
\end{figure}

Figure~\ref{fig:hf_phasediagrams} demonstrates the effect of local chemistry by overlaying the phase diagrams obtained for the cubic-perovskite Sr and La materials. We see that when expressed in the $U$-$N_d$ plane there is almost no difference in the location of the metal-insulator phase boundary, except in the region of $N_d$ very near the atomic limit where small differences in the (very small, but not zero) $d$-$d$ hopping lead to slight differences in the location of the phase boundaries. We conclude that for the hypothetical cubic structures the only difference between the Sr and La materials is the different electronegativities of the transition metal ions, which lead to differences in the $p$-$d$ energy splitting and thus to the $d$ occupancies of the transition metal ions. This again underscores the importance of charge transfer physics in  early transition metal compounds. 

The phase diagrams for LaTiO$_3$ and LaVO$_3$  in  the experimental  (GdFeO$_3$-distorted) structure are  shown in Fig.~\ref{fig:s_l_comp}. The unit cell of the GdFeO$_3$-distorted structure contains four transition-metal ions; thus ``staggered'' (in cubic notation) phases may be found in a DFT+Hartree calculation even without further spatial symmetry breaking. However, to understand the effect of the lattice distortion in this figure we restrict our study to ferromagnetic and ferro-orbital states (in cubic notation), in other words we require that the spin and orbital states of each octahedron are the same. 

\begin{figure}[t]
    \centering
    \includegraphics[width=\columnwidth]{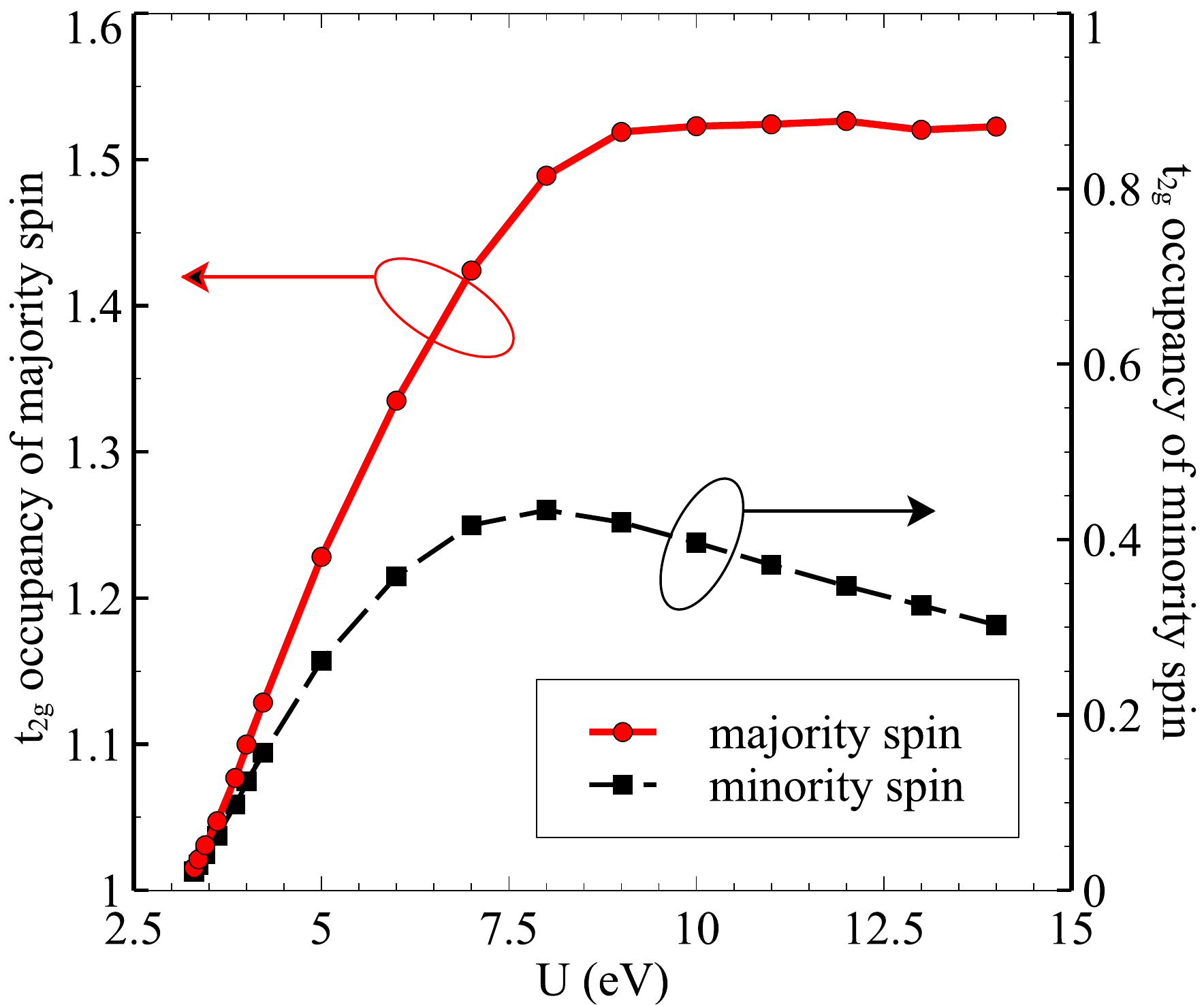}
    \caption{\label{fig:hf_svo_docc_spinresolved}(Color online) The $d$ occupancy of majority and minority spins for SrVO$_3$ as a function of $U$ with the total $N_d$ chosen so that the system is at the MIT (going along the SrVO$_3$ phase boundary - see Fig.~\ref{fig:s_l_comp}a).}
\end{figure}

Figure~\ref{fig:s_l_comp} shows that the main effect of the GdFeO$_3$ distortion is to shift the location of the phase boundaries: the insulating state extends over a wider parameter range in the GdFeO$_3$-distorted structure than in the cubic structure, and the enhancement is larger for LaTiO$_3$ than for LaVO$_3$. One might imagine that a significant contribution to the difference arises from the decrease in bandwidth caused by the  GdFeO$_3$ distortion. In the distorted structure, the $B$-O-$B$ bond ($B$ is the transition metal atom) is buckled, reducing the amplitude for an electron to hop from one $B$-site to the next. In our DFT calculations (not shown), the bandwidth of the $p$-$d$ antibonding bands is reduced by $25\%$ for LaVO$_3$ and $\sim20\%$ for LaTiO$_3$. However, we observe that a decrease in bandwidth is equivalent to an increase of $U$, in other words, to a vertical shift of the phase boundary in Fig.~\ref{fig:s_l_comp}. As can be seen by inspection of the figures, vertically shifting the curve obtained for the cubic system does not make it coincide with the phase boundary obtained for the distorted one.  In fact, as already noted by Pavarini and co-workers \cite{Pavarini04}, the key physics is that the distortion breaks the orbital symmetry, leading to a ``1 down 2 up'' distortion that promotes orbital order. This orbital order strongly favors the insulating state in LaTiO$_3$. However, in high-spin  $d^2$ systems such as LaVO$_3$ the natural symmetry breaking associated with an insulating phase would be of the ``2 down 1 up'' type, which is not produced by the GdFeO$_3$ distortion. The  actual ``1 down 2 up'' orbital splitting  has a much smaller effect.   

Figure~\ref{fig:s_l_comp} also shows that at very large $U$ the phase boundary is not vertical, but bends back. The back-bending  reflects the decrease of the occupancy of high-lying minority spin $d$ states as they are pushed to very high energies by the large $U$. Calculations (Fig.~\ref{fig:hf_svo_docc_spinresolved}) of the spin-resolved $d$ occupancy show that, in the $d^1$, $d^2$ and $d^3$ cases, the majority spin $d$ occupancy tends to a $U$-independent asymptote as $U$ is increased while the minority spin occupancy decreases. We expect that as $U\rightarrow\infty$, the minority spin $d$ occupancy goes to zero and the phase boundary in the $U$-$N_d$ plane asymptotes to a vertical ($U$-independent) line. For GdFeO$_3$-distorted cases, the bending of LaTiO$_3$ is larger than that of LaVO$_3$ (see Fig.~\ref{fig:s_l_comp}) because of the splitting of unoccupied orbitals arising from the  strong orbital polarization. This strong orbital polarization allows a decrease in  the occupancy of the nominally empty majority spin orbitals in addition to the decrease in minority spin occupancy. In LaVO$_3$ ($d^2$ systems), the orbital splitting is negligible and there is no such effect.  We note that this back-bending is amplified in the DFT+Hartree calculations by  the strong spin and orbital polarization found in this approximation. In the DMFT calculation, as shown later, the unoccupied states are not split as much and so this behavior is less pronounced. 

The DFT+Hartree method combined with the standard FLL double counting, a physically reasonable value of $U\sim 5$eV and the experimental structure predicts an  insulating state for LaTiO$_3$ and LaVO$_3$ in agreement with experiment.  We believe that this apparent agreement arises from a cancellation of errors and is simply fortuitous.  The two errors are that Hartree approximations are known to overestimate order and therefore favor insulating states, and that the DFT approximation overestimates $N_d$ and therefore underestimates the tendency to order. Later in the manuscript we will demonstrate that the same calculation using DFT+DMFT(QMC) results in a metallic state, and we show that the double-counting must be adjusted in order to properly capture the insulating state and the spectra. 

\section{DFT+DMFT calculations\label{sec:dmft}}

The Hartree approximation does not include quantum fluctuations arising from electronic correlations and cannot capture the paramagnetic Mott insulating phase, which is observed experimentally in many early transition-metal oxides \cite{Imada98} including LaTiO$_3$ and LaVO$_3$. To treat the correlation more properly, it is necessary to go beyond the mean field approximation. In this section, we use the dynamical mean-field method to study the metal-insulator transition in the paramagnetic DFT+DMFT framework \cite{Kotliar06}.

The procedure follows the one discussed above in the context of the Hartree approximation, but with the local self-energy computed using the single-site DMFT approximation rather than the Hartree approximation. A DMFT solution for the full five orbital model is too expensive for wide surveys of parameter space. In most of our DMFT calculations, we therefore use the model with transition metal $t_{2g}$ bands (and oxygen $p$ bands), but for selected points we present results obtained with the full five-orbital model.

\begin{figure}[t]
 \centering
 \includegraphics[width=\columnwidth]{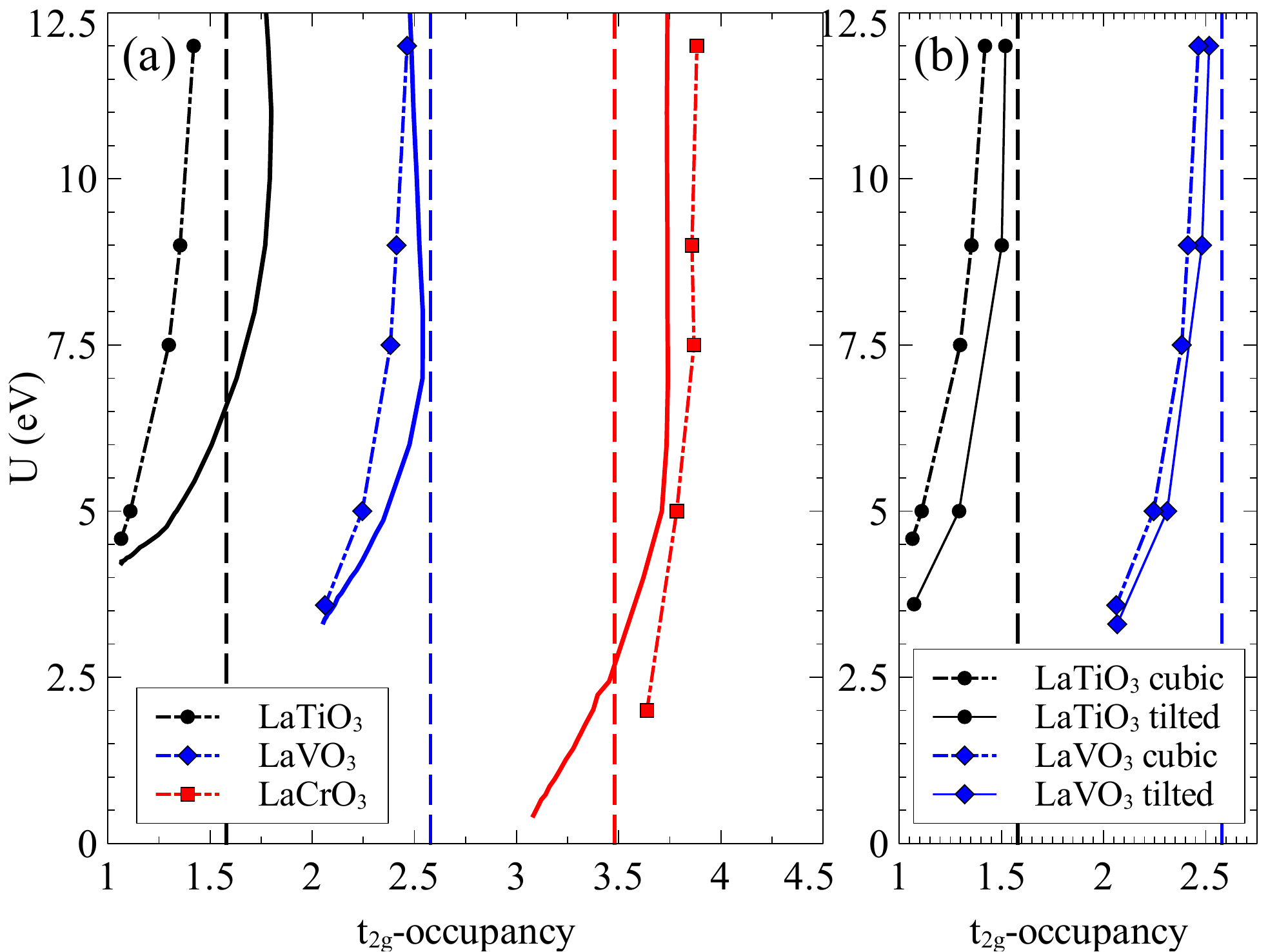}
\caption{\label{fig:dmft_la_series}(Color online) The  metal insulator phase diagrams of LaTiO$_3$, LaVO$_3$ and LaCrO$_3$ calculated using the DFT+DMFT and DFT+Hartree procedures described in the text but retaining only the $t_{2g}$ portion of the $d$ manifold,   presented  in the $U$-$N_d$ plane. (a) Metal-insulator phase diagrams calculated  assuming that the materials are in the cubic perovskite structure   using DFT+DMFT (dashed-dotted curves with symbols) and DFT+Hartree  (solid curves). (b) The phase boundaries obtained using the experimental (GdFeO$_3$-distorted) structures using DFT+DMFT. The vertical dashed lines are $t_{2g}$ occupancies derived from DFT+MLWF. The temperature is $T=0.1$eV. The insulating (metallic) regime is to the left (right) of the phase boundary. }
\end{figure}

We first compare the results  obtained using DMFT and Hartree calculations for the La series in Fig.~\ref{fig:dmft_la_series}. Panel (a) shows results obtained for a hypothetical cubic structure. The difference between the DMFT and Hartree phase boundaries  is substantial in LaTiO$_3$ and much less for the other two  compounds.  As the nominal number of $d$ electrons increases, the DMFT solution becomes more insulating and in the $d^3$ case (LaCrO$_3$), DMFT predicts a larger insulating regime than does the Hartree calculation.

This change in DMFT phase boundary is a consequence of the Hund's coupling $J$, which behaves differently for systems with different $d$ valence. Some aspects of the differences between materials can be understood from atomic-limit estimates, following Ref.~\cite{Georges13}. In the atomic limit the energy cost to move a $d$ electron from one transition metal atom with $N$ valence electrons to another is $\Delta(N) = E(N+1)+E(N-1)-2E(N)$. At half-filling ($N=3$), $\Delta(N=3)=U+2J$, while for $d^1$ and $d^2$ cases, $\Delta=U-3J$. Therefore, with $J\ne0$, we expect $d^3$ systems have the largest gaps and hence the smallest $U_c$ for the MIT, explaining the large enhancement of insulating regime in LaCrO$_3$. However, the atomic limit gaps for atomic $d^1$ and $d^2$ are equal, suggesting that  LaTiO$_3$ and LaVO$_3$ should have comparable $U_c$. The differences in phase boundary arise because in the vanadate case there is some admixture of $d^3$ into the ground state, leading to more insulating behavior. This difference is thus a consequence of charge transfer physics in the early transition-metal oxides. 

It is interesting to note that for LaCrO$_3$  the DMFT calculation has a larger regime of insulating behavior than the Hartree calculation. This does not contradict the general statement that the Hartree approximation overestimates order; it merely shows that our DFT+Hartree calculations, which were restricted to ferromagnetic and ferro-orbital states, did not include the correct long-ranged order for this compound. Calculations (not shown) allowing for antiferromagnetic order would produce a much larger regime of insulating behavior.

\begin{figure}[t]
    \centering
    \includegraphics[width=\columnwidth]{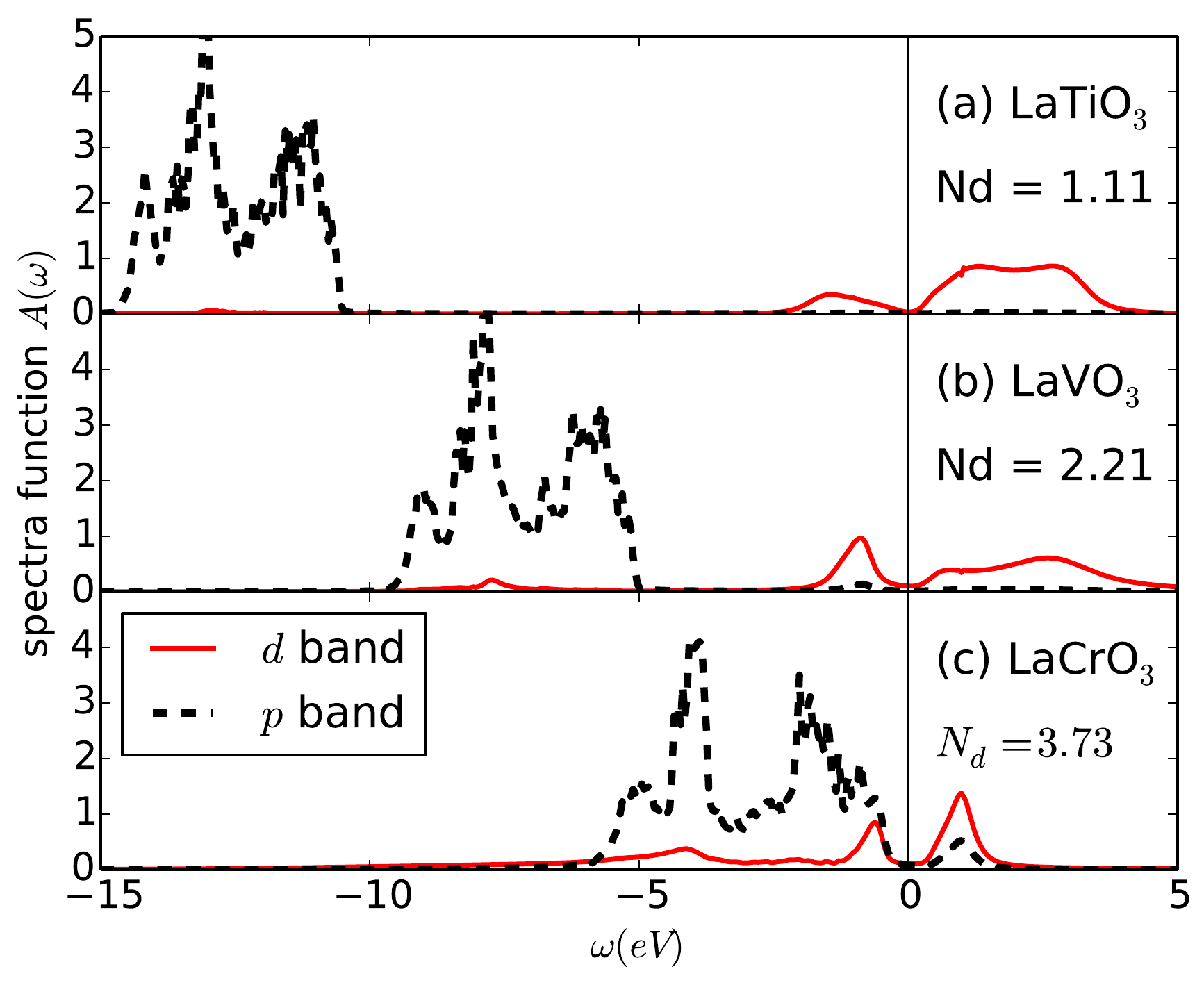}
    \caption{\label{fig:spec_cubic_la}(Color online) Spectral functions $A(\omega)$ for cubic LaTiO$_3$, LaVO$_3$ and LaCrO$_3$ at $U=5$eV, $J=0.65$eV and $N_d$ chosen to be close to the MIT phase boundaries. The dashed curves (black online) are oxygen $p$ bands, the solid curves (red online) are correlated $d$ bands. The vertical line marks the Fermi level.}
\end{figure}

Figure~\ref{fig:spec_cubic_la} shows the one-electron spectral functions (interacting DOS) of hypothetical cubic  LaTiO$_3$, LaVO$_3$ and LaCrO$_3$ with $U=5$eV, which is around the typical $U$ value computed for early transition-metal oxides in the  perovskite structure \cite{Vaugier12}. $N_d$ is adjusted so that the systems are insulating but close to the transition to the metallic state. We see that to drive hypothetical cubic  LaTiO$_3$ into the insulating state one must shift the  oxygen bands to about $-10$eV far from the Fermi level. This energy for the oxygen states is in very substantial disagreement with the experimental value $\sim -3.5$eV although the gap between the highest occupied states and the lowest unoccupied ones is consistent with the experimental value.  For LaVO$_3$ an insulating state can be obtained for  oxygen bands closer ($-5$eV) to the Fermi level, but this oxygen band energy remains in substantial disagreement with experiment. Finally, in LaCrO$_3$ an insulating state can be obtained even for $p$ states very close to the Fermi level. These observations illustrate the necessity of including the octahedral rotations.

We now discuss the results of DMFT calculations for LaTiO$_3$ and LaVO$_3$ using the experimental  structures. Some of the results have been partly discussed in Ref.~\cite{Dang13}. Here, we go beyond the results of Ref.~\cite{Dang13}, in particular discussing in detail the effects of the structural distortion and providing a comparison to the DFT+Hartree calculations. The MIT phase diagrams for LaTiO$_3$ and LaVO$_3$ obtained using DMFT calculations performed using the experimental  structures (solid curves) are shown in Fig.~\ref{fig:dmft_la_series}b in comparison with those obtained  using the hypothetical cubic structure (dotted-dashed curves). As in the DFT+Hartree calculations, increasing the magnitude of the rotational lattice distortion enlarges the insulating regime, both in $U$ and $N_d$ with the increase being larger in LaTiO$_3$ than in LaVO$_3$.

\begin{figure}[t]
    \centering
    \includegraphics[width=\columnwidth]{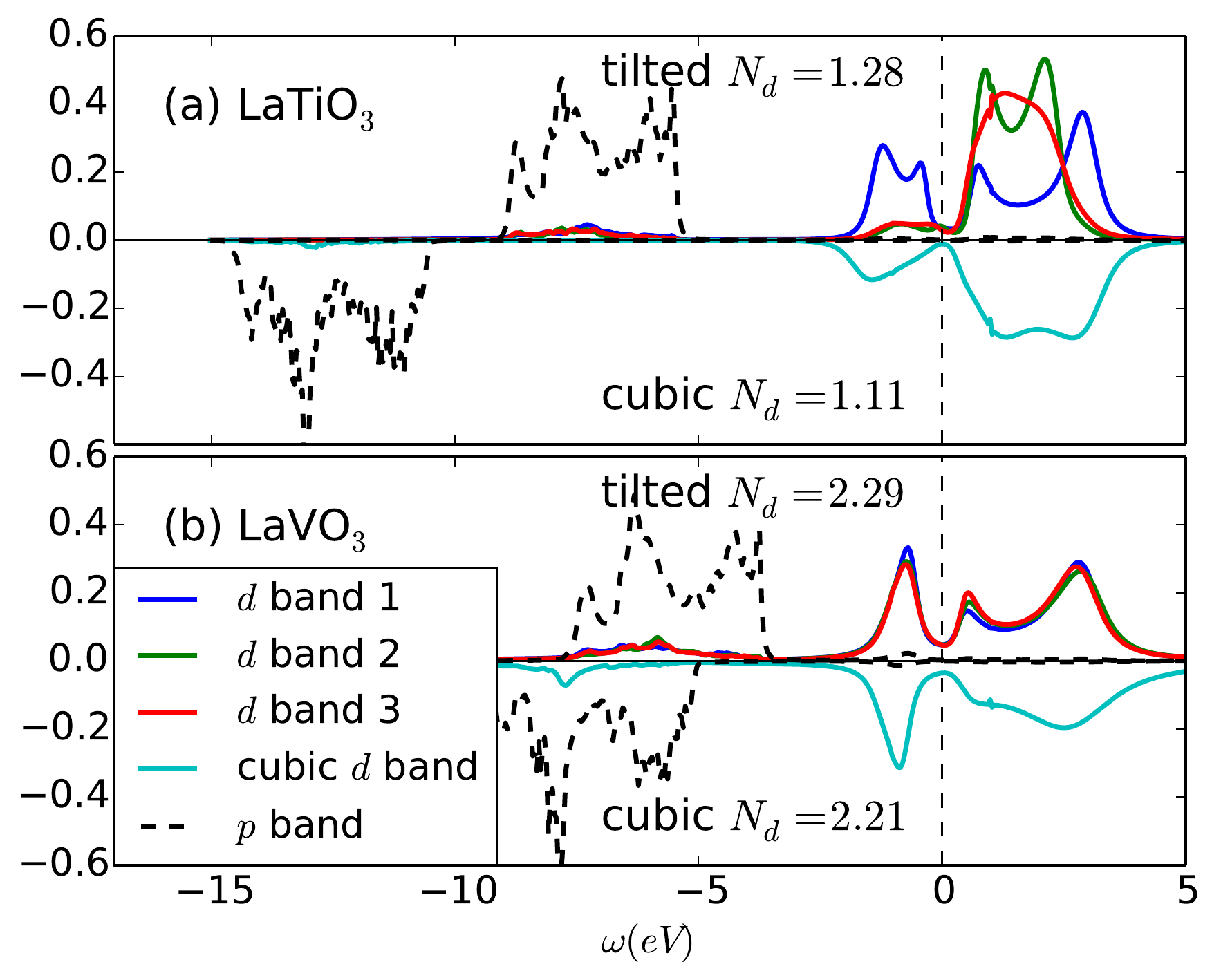}
    \caption{\label{fig:spec_tilted_la}(Color online) Spectral functions $A(\omega)$ for cubic (negative value) and GdFeO$_3$-distorted (positive value) structures of LaTiO$_3$ and LaVO$_3$ at values $U=5$eV, $J=0.65$eV and $N_d$ chosen to be close to the MIT phase boundaries. The dashed curves (black online) are the average spectra per band for oxygen $p$ bands, the solid curves (color online) are correlated $d$ bands. The vertical dashed line marks the Fermi level.}
\end{figure}

The qualitative similarity of the DMFT and Hartree results  (see Sec.~\ref{sec:hartree}) suggests that insights gained from the Hartree calculations can be applied to understand the DMFT results.  First, the lattice distortion decreases the antibonding bandwidth $W$ by $20\to25\%$. In the Mott-insulating regime (small $U$ region), the critical $U_c$ for Mott transition is proportional to $W$, with a smaller bandwidth in the distorted structure, so the critical  $U$ becomes smaller. In the charge transfer regime (large $U$ region), the reduced $p$-$d$ hybridization  means oxygen $p$ bands must come closer to the $d$ states in order to induce enough covalency to destroy the insulating state. The difference in the enhancement of the insulating regime between LaTiO$_3$ and LaVO$_3$ arises from  orbital ordering. In the experiment structure, the $t_{2g}$ orbitals in both materials experience a crystal field splitting. In the DFT calculation, LaTiO$_3$ has a weak ``1 up, 2 down'' orbital order (one orbital is occupied more than the other two). The DMFT results indicate that  this type of orbital ordering is enhanced significantly by interactions [Fig.~\ref{fig:spec_tilted_la}a], although the precise degree of enhancement depends on the value of $N_d$. For LaVO$_3$, the DMFT calculation indicates that there is almost no orbital order. We believe that the lack of orbital ordering occurs because the virtual charge fluctuations in the $d^2$ state lead to a significant admixture of $d^3$, and the tendency of the  Hund's coupling to favor high spin then reduces the tendency to order.  Thus, with the DMFT approximation we conclude that in LaTiO$_3$ the orbital splitting induced by the  GdFeO$_3$ rotation is essential for Mott behavior, while in LaVO$_3$, the main effect of the distortion is to reduce the bandwidth. The bandwidth reduction  also favors  order, but to a lesser degree \cite{Werner09,Georges13}.

\section{Determining Physicially Relevant Values for $U$ and the Double Counting\label{sec:positions}}

\begin{figure}[t]
    \centering
    \includegraphics[width=\columnwidth]{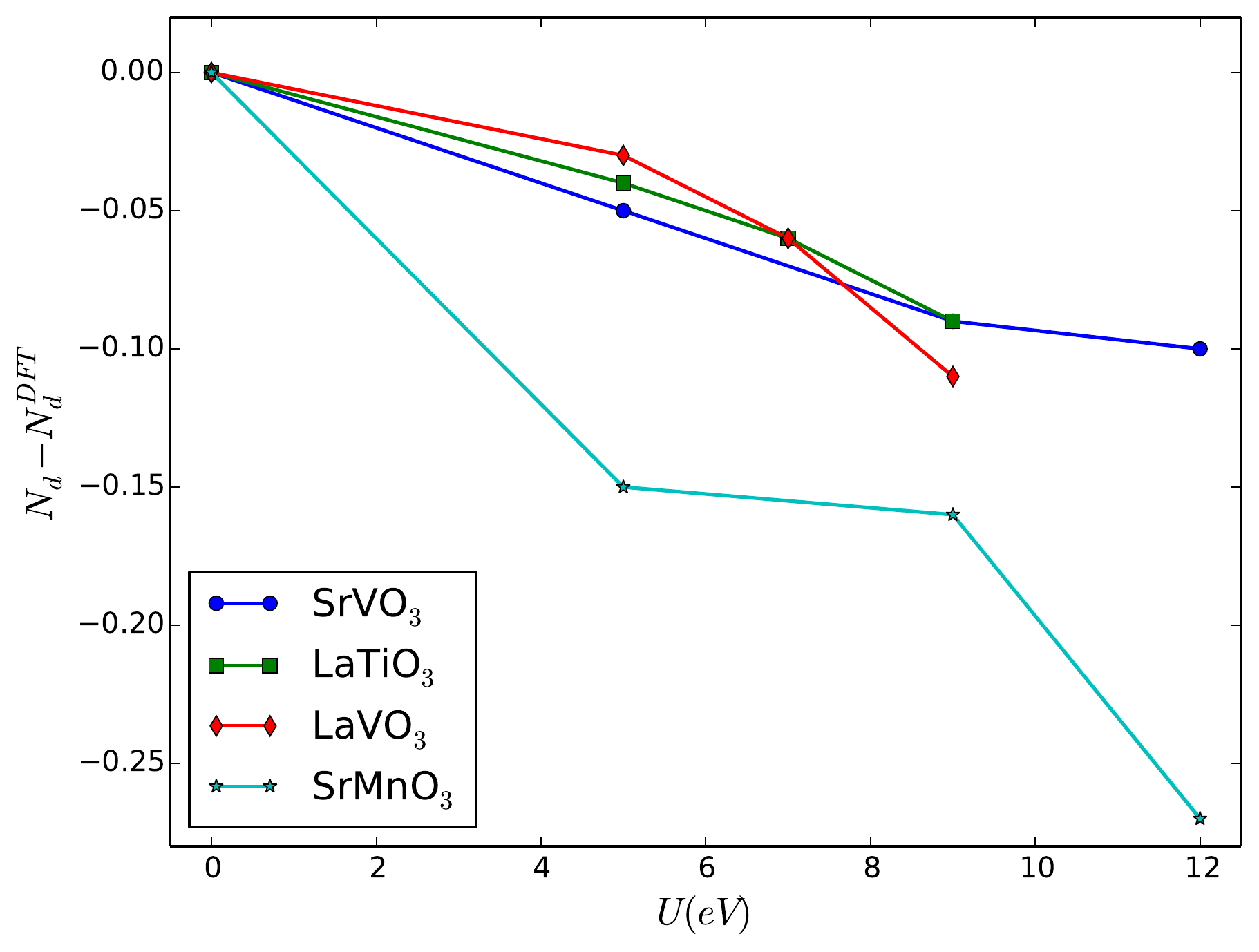}
    
      \caption{\label{fig:full_charge_nd}(Color online) The dependence of $d$ occupancy $N_d$ on the interaction $U$ using fully charge self consistent calculations with FLL double-counting correction (Wien2k+TRIQS code \cite{Aichhorn11,triqs_project}). The temperature is $T=0.1$eV. The calculations use projector method \cite{Aichhorn09} to obtain the full five correlated $d$ orbitals. Note that the $y$ axis is the difference between $N_d$ and the DFT value $N_d^{DFT}$ where the $N_d^{DFT}$ values are $2.60,1.81,2.92$ and $4.81$ for SrVO$_3$, LaTiO$_3$, LaVO$_3$ and SrMnO$_3$, respectively.}
\end{figure}

In previous sections, we studied the general structure of the theoretical results, varying the $p$-$d$ splitting and interaction strength over wide ranges. In this section we ask how to choose reasonable values for  the actual systems by estimating the interaction and $p$-$d$ splitting parameters.

\begin{figure*}[t]
    \centering
    \includegraphics[width=0.42\textwidth]{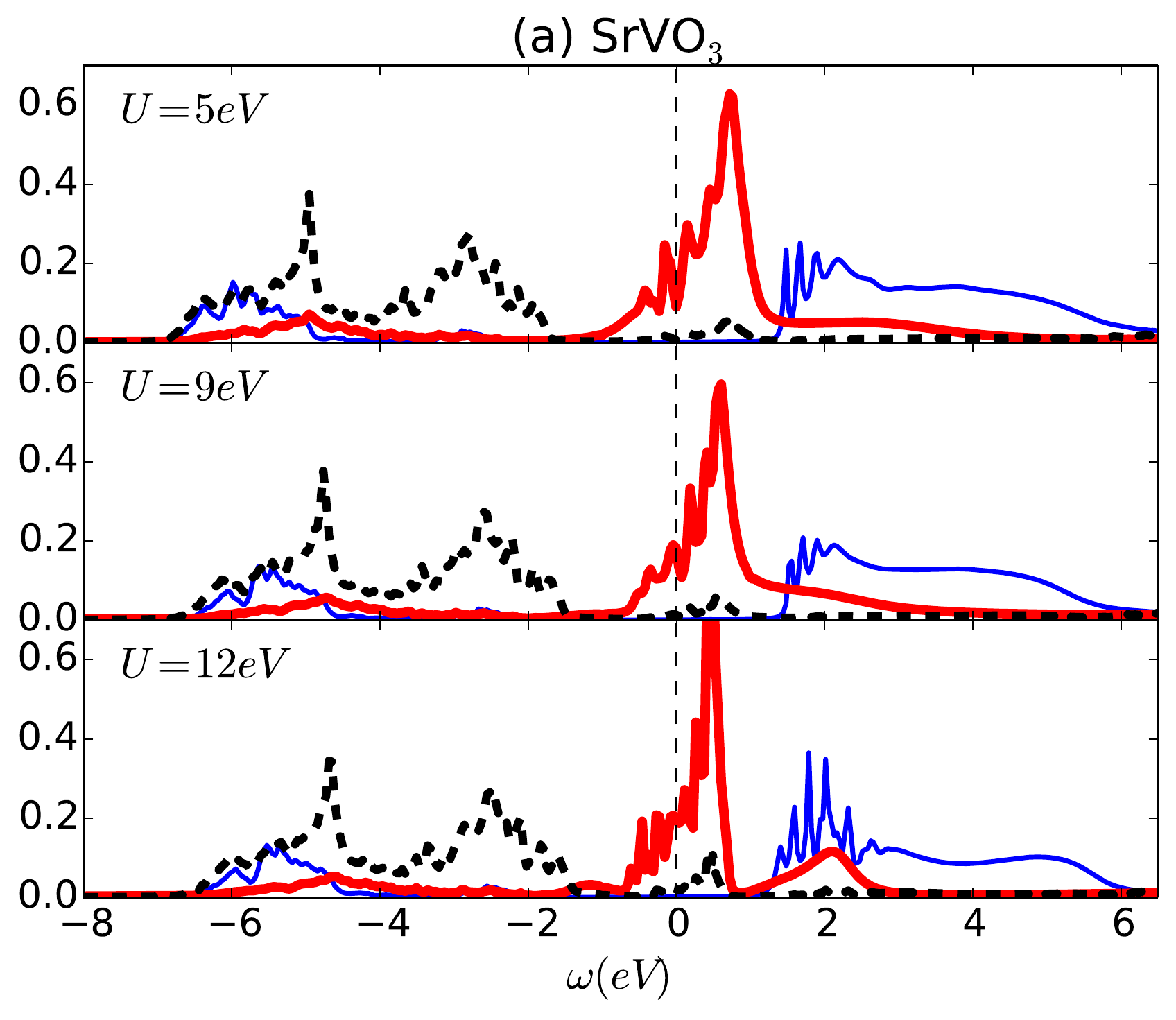}
    \includegraphics[width=0.42\textwidth]{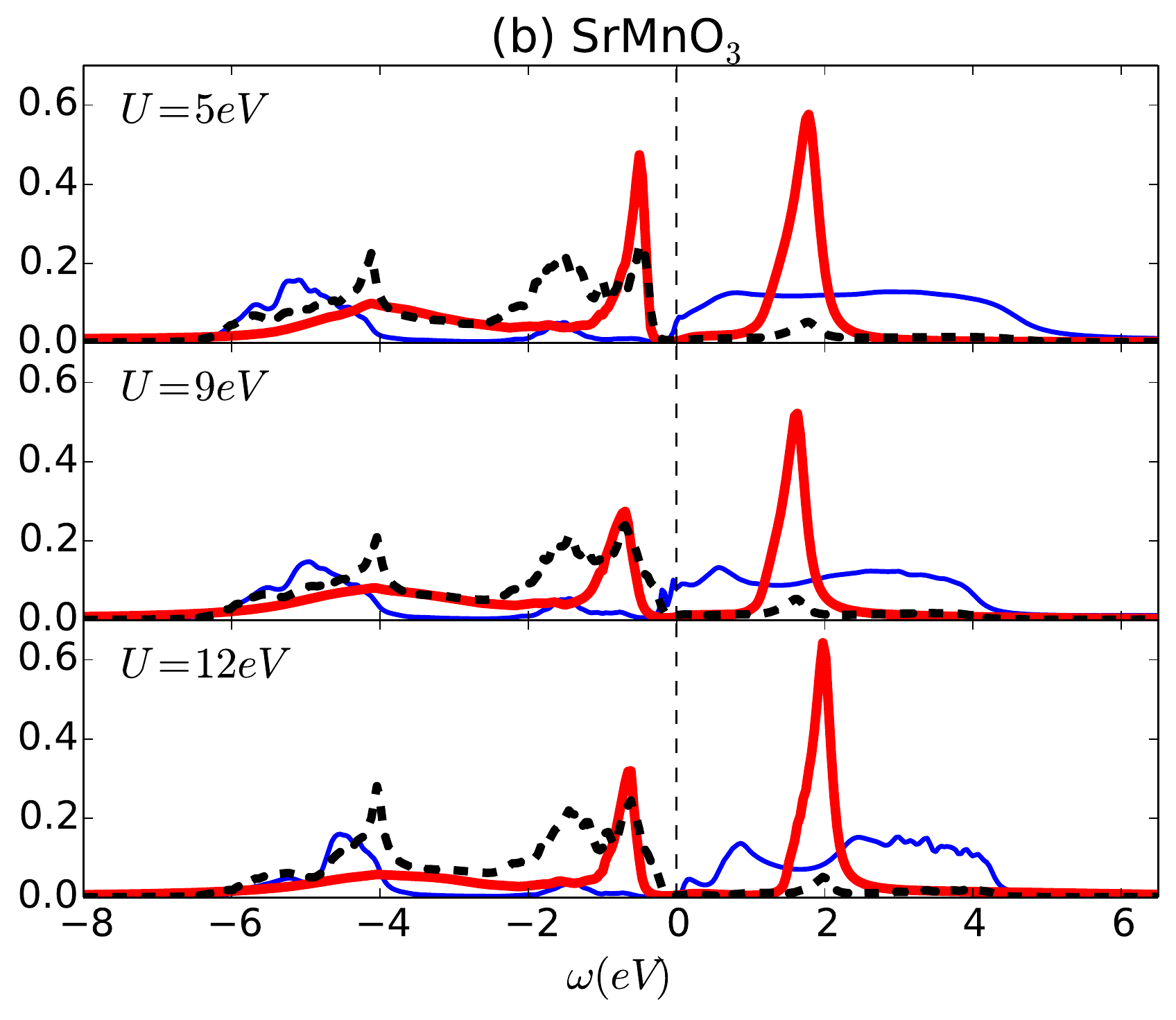}
    
    \includegraphics[width=0.42\textwidth]{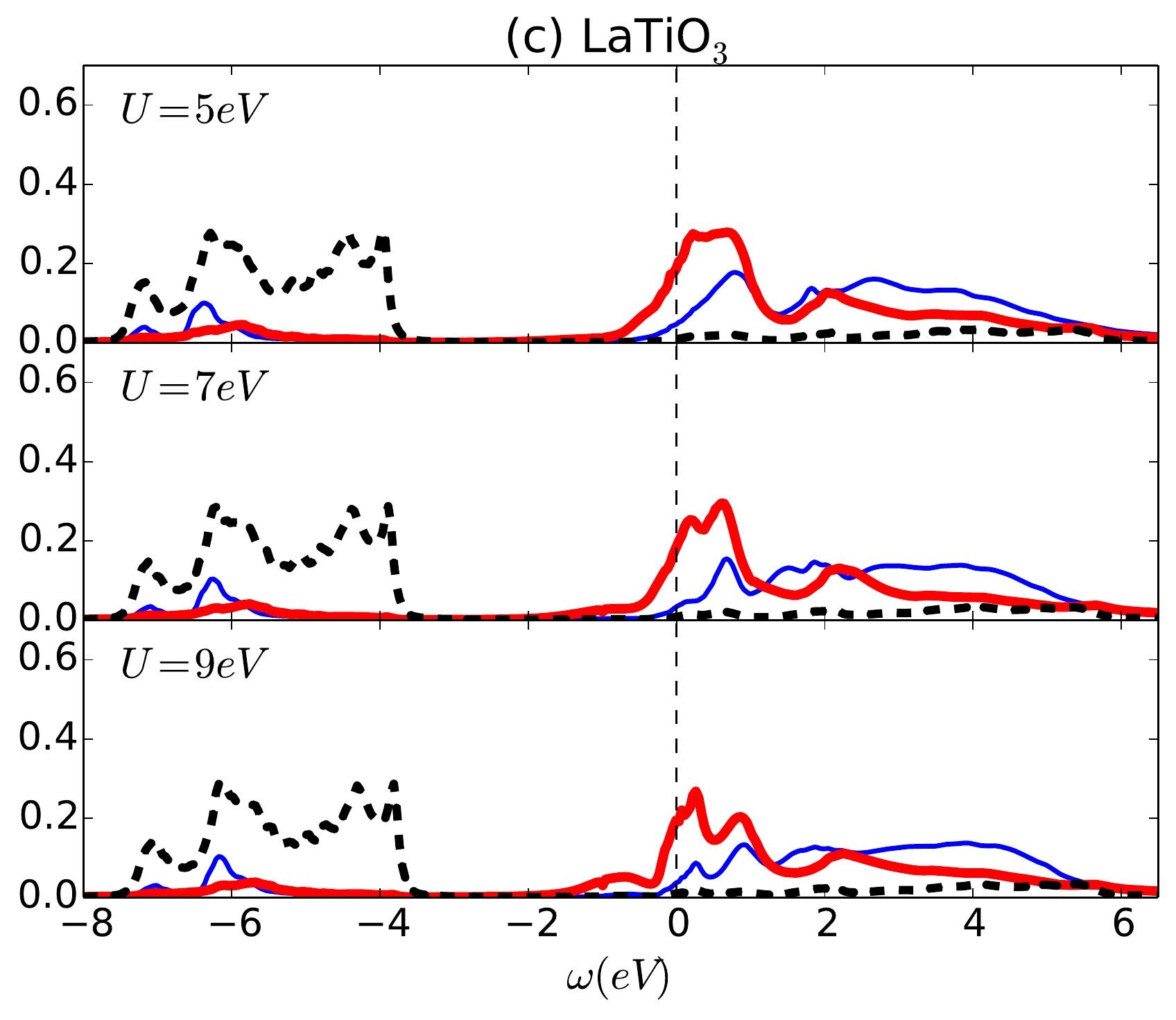}
    \includegraphics[width=0.42\textwidth]{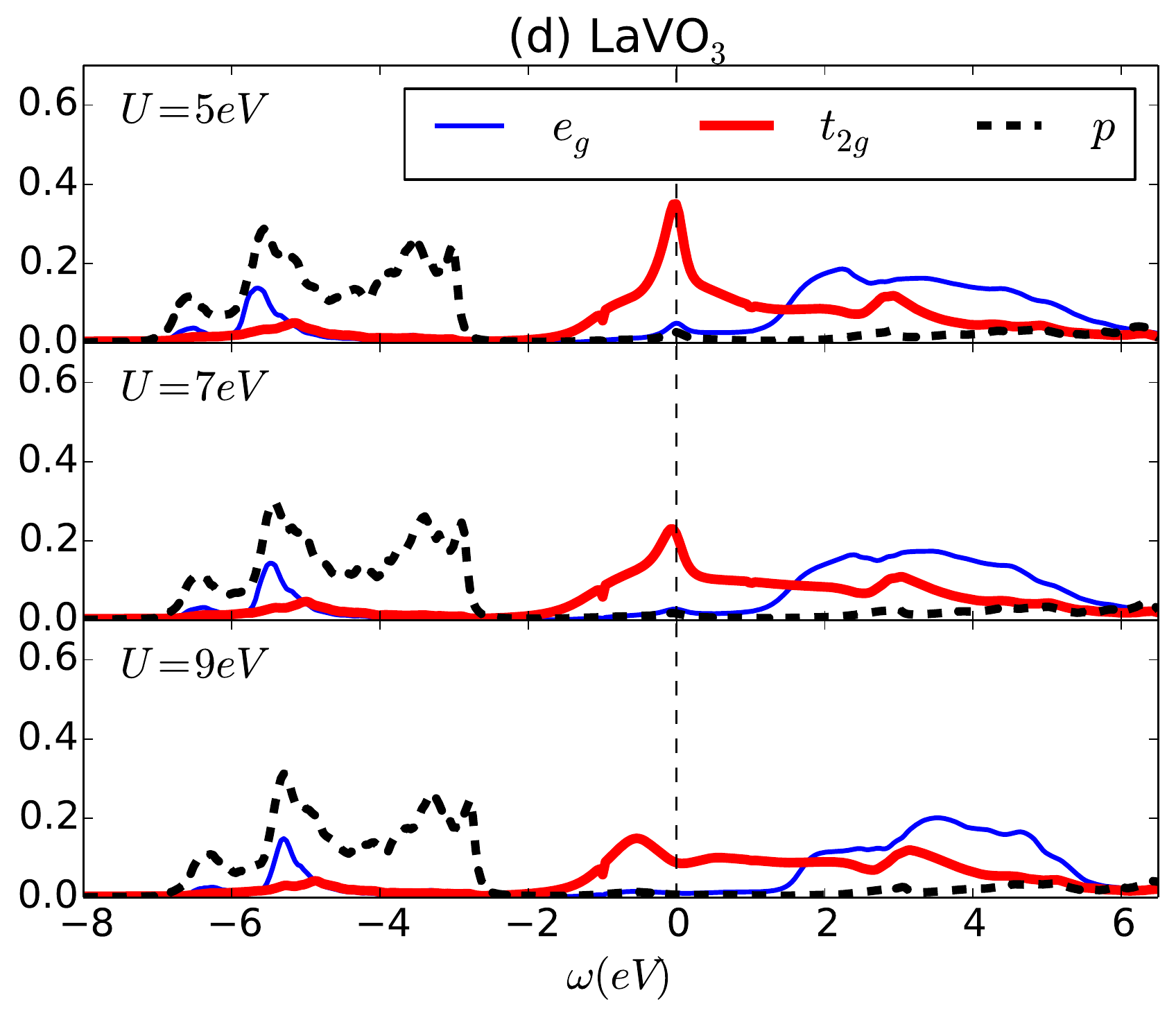}
    \caption{\label{fig:full_charge_spectra}(Color online) Full charge self consistent spectral functions using FLL double-counting correction (Wien2k/TRIQS) for (a) SrVO$_3$, (b) SrMnO$_3$, (c) LaTiO$_3$, and (d) LaVO$_3$ at various $U$ values. The spectra are corresponding to the $N_d$ versus $U$ plots in Fig.~\ref{fig:full_charge_nd} at $U\ne 0$. Note that only the average for each $t_{2g},e_g$ and $p$ types is plotted in order to make the plots easy to see.}
\end{figure*}

First, we specify the correct values of the Hubbard value $U$ and the Hund's coupling $J$ [defined in Eq.~\eqref{eq:onsite_SlaterKanamori}] for materials. The Hund's coupling is only weakly renormalized by solid state effects \cite{Aryasetiawan06,Vaugier12}, and is believed to be of the  order of $1$eV or slightly less.  Ref.~\cite{Vaugier12} shows that $J$ is around $0.65$eV for SrVO$_3$, SrCrO$_3$ or SrMnO$_3$ (using the energy window including $p$ and $d$ bands and symmetrizing over the interactions of the $t_{2g}$ bands) and we adopt this throughout our paper.  In contrast, the $U$ value is screened strongly \cite{Aryasetiawan06,Vaugier12}, being  five or six times smaller than the bare value, with the precise renormalization depending on material parameters. For SrVO$_3$, Ref.~\cite{Vaugier12} estimates   $U=4.1$eV (note we have expressed the result of Ref.~\cite{Vaugier12}  using the Kanamori parametrization). Because the La-based materials are Mott insulators one might expect the screening to be slightly less, so the $U$ values might correspondingly be slightly larger.   In Ref.~\cite{Dang13}, we show that within the MLWF scheme, only a range of $U\sim 6\pm 1$eV can reproduce both the observed insulating gap and the position of the oxygen states so we suggest that this value is reasonable. We note however that our results are not strongly sensitive to $U$.

We also note that a recent paper \cite{Haule13} using fully-charge self consistent DFT+DMFT calculations with $d$ states defined via a projector method argued that $U=10$eV is reasonable for oxides.  The origin of this difference requires further investigation. One important issue is the difference in $p$-$d$ hybridization between Wannier and projector methods. The relatively larger $p$-$d$ hybridization in the projector method requires a larger $U$ to obtain an insulating state. Other technical differences occur in the calculation, including in particular the use of a broader energy window, incorporating e.g. La-derived bands.  A calculation of the screened Coulomb interaction within the system defined in Ref.~\cite{Haule13} would be of interest. 

\begin{table}[t]
\begin{ruledtabular}
    \begin{tabular}{ccccc}
                                        & SrVO$_3$ & LaTiO$_3$ & YTiO$_3$ & LaVO$_3$ \\ \hline 
        exp. energy gap               & 0       & 0.3eV     & 1eV      & 1eV \\ 
        exp. oxygen bands position & 2.4eV & 5.35eV & 4.95eV   & 4.35eV \\
        DFT oxygen bands position & 1.5eV & 3.25eV & 3.15eV   & 2.5eV \\ 
        exp. $N_d$               & 1.73 & 1.28 & 1.31 & 2.24 \\
        DFT $N_d$                & 1.99 & 1.56 & 1.57 & 2.55 \\
        
    \end{tabular} 
    \end{ruledtabular}
    \caption{\label{table:exp_data} The first row is experimental data for the energy gaps from Ref.~\cite{Arima93}. The values of ``exp. oxygen bands position'' and ``exp. $N_d$'' are the $p$ band positions and the $d$ occupancy values obtained from Fig.~\ref{fig:spec_with_exp} where the spectra match experiments. The ``DFT $N_d$'' values are from DFT calculations (with MLWF method) and the ``DFT oxygen band positions'' are the $p$ band positions obtained from Fig.~\ref{fig:spec_dftnd}.}
\end{table}

The next crucial issue is the value of the $p$-$d$ energy difference or double-counting correction, parametrized here by the $d$ occupancy $N_d$. We first note that the $N_d$ values can be different depending on the method used to define the correlated subspace. The results presented in this section are obtained using one of three different methods: VASP projector (in DFT+$U$ calculations), Wien2k+TRIQS projector (in fully-charge self consistent calculations) and MLWF method (in ``one-shot'' DMFT calculations). We will therefore note the method together with the $N_d$ value. 

We carried out fully charge self consistent calculations using the DFT+DMFT framework with realistic structures and FLL double counting. Figure~\ref{fig:full_charge_nd} shows $N_d-N_d^{DFT}$ as a function of $U$ for several materials. The changes are small relative to the total $N_d$  for $d^1$ and $d^2$ materials (SrVO$_3$, LaTiO$_3$, LaVO$_3$). Figures~\ref{fig:full_charge_spectra}13(a),13(c),13(d) are the spectra of SrVO$_3$, LaTiO$_3$ and LaVO$_3$ corresponding to the $U$ values used in Fig.~\ref{fig:full_charge_nd}. As the $d$ occupancies of these materials do not change much, the change in the spectra of these materials are insignificant.  More importantly, the DFT+DMFT calculations with full charge self consistency and FLL double counting predict that all of the $d^1$ and $d^2$ perovskites are  metals as can be seen directly from the calculated spectra shown in Fig.~\ref{fig:full_charge_spectra}. We conclude that the standard DFT+DMFT with FLL double counting does not put materials in the correction positions in the phase diagrams, as from experiments, LaTiO$_3$ and LaVO$_3$ are  Mott insulators \cite{Arima93}.

\begin{figure}[t]
    \centering
    \includegraphics[width=\columnwidth]{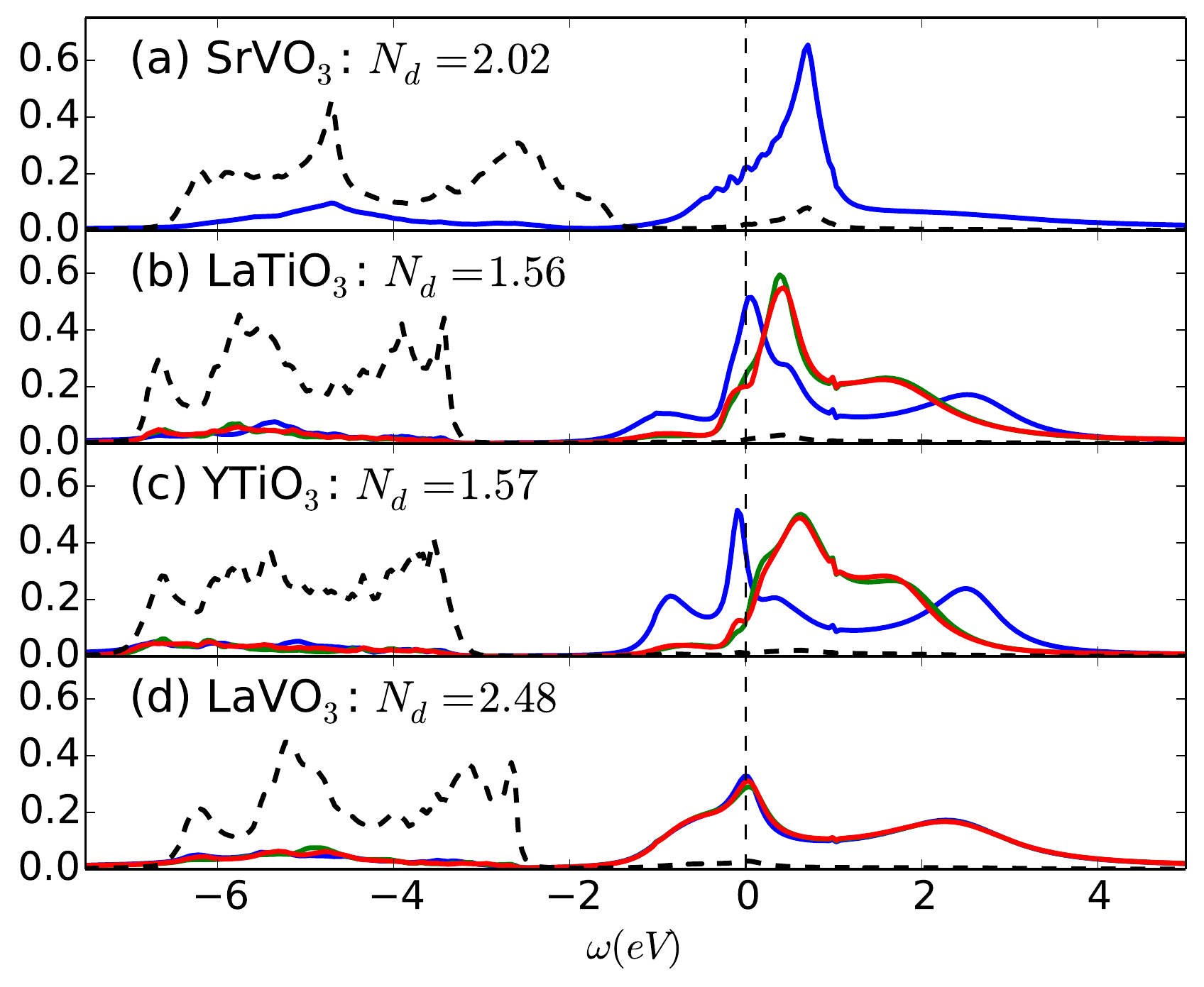}
    \caption{\label{fig:spec_dftnd}(Color online) Spectral functions $A(\omega)$ for SrVO$_3$, LaTiO$_3$, YTiO$_3$ and LaVO$_3$ using realistic lattice structure at $U=5$eV, $J=0.65$eV and $N_d$ (obtained from MLWF method) chosen to be close to the value from \textit{ab initio} calculations. The dashed curves (black online) are the average spectra per band for oxygen $p$ bands, the solid curves (color online) are the three correlated $t_{2g}$ bands. The vertical dashed line marks the Fermi level.}
\end{figure}

The case of SrMnO$_3$ is different. When applied to SrMnO$_3$ the fully charge self-consistent DMFT procedure leads to an $N_d$ significantly smaller than the DFT value (see Fig.~\ref{fig:full_charge_nd}) and places the material at the edge of the insulating regime. Part of the difference from the $d^1$ and $d^2$ materials may relate to the half-filled nature of the $t_{2g}$ shell, but understanding why the Mn material is so different from the others remains an important open problem.

As shown above, the fully charge self consistent results yield $d$ occupancies for $d^1$ and $d^2$ systems that are close to the DFT values. We thus conduct ``one-shot'' DMFT calculations (using the MLWF correlated subspace) for SrVO$_3$, LaTiO$_3$, LaVO$_3$ and YTiO$_3$ with the double-counting correction adjusted to have the $d$ occupancies close to the DFT values. This will elucidate the role of using a different type of correlated subspace. Figure~\ref{fig:spec_dftnd} shows the spectra at $U=5$eV, in which all materials are in metallic state, confirming that our results are not dependent on the details of the correlated subspace. Moreover, the two different methods used to produce  Fig.~\ref{fig:full_charge_spectra} (the projector method \cite{Aichhorn09}) and Fig.~\ref{fig:spec_dftnd} (MLWF method \cite{PhysRevB.56.12847,PhysRevB.65.035109}) give oxygen $p$ bands positions quite close to each other, with the largest difference found in  LaTiO$_3$ where the difference in the $p$ band position is about $0.7$eV. Even though there are differences between different projection methods, the spectra in Figs.~\ref{fig:full_charge_spectra},\ref{fig:spec_dftnd} show that full charge self consistency is unnecessary: if the $d$ occupancy is known, one can reproduce the fully charge self consistent result using one-shot calculation with the $N_d$ adjusted to the known value. Of course, this presumes that one is using a normalized projector which is defined over a similar energy region as the Wannier functions which are used to define
the correlated subspace.

\begin{figure}[t]
    \centering
    \includegraphics[width=\columnwidth]{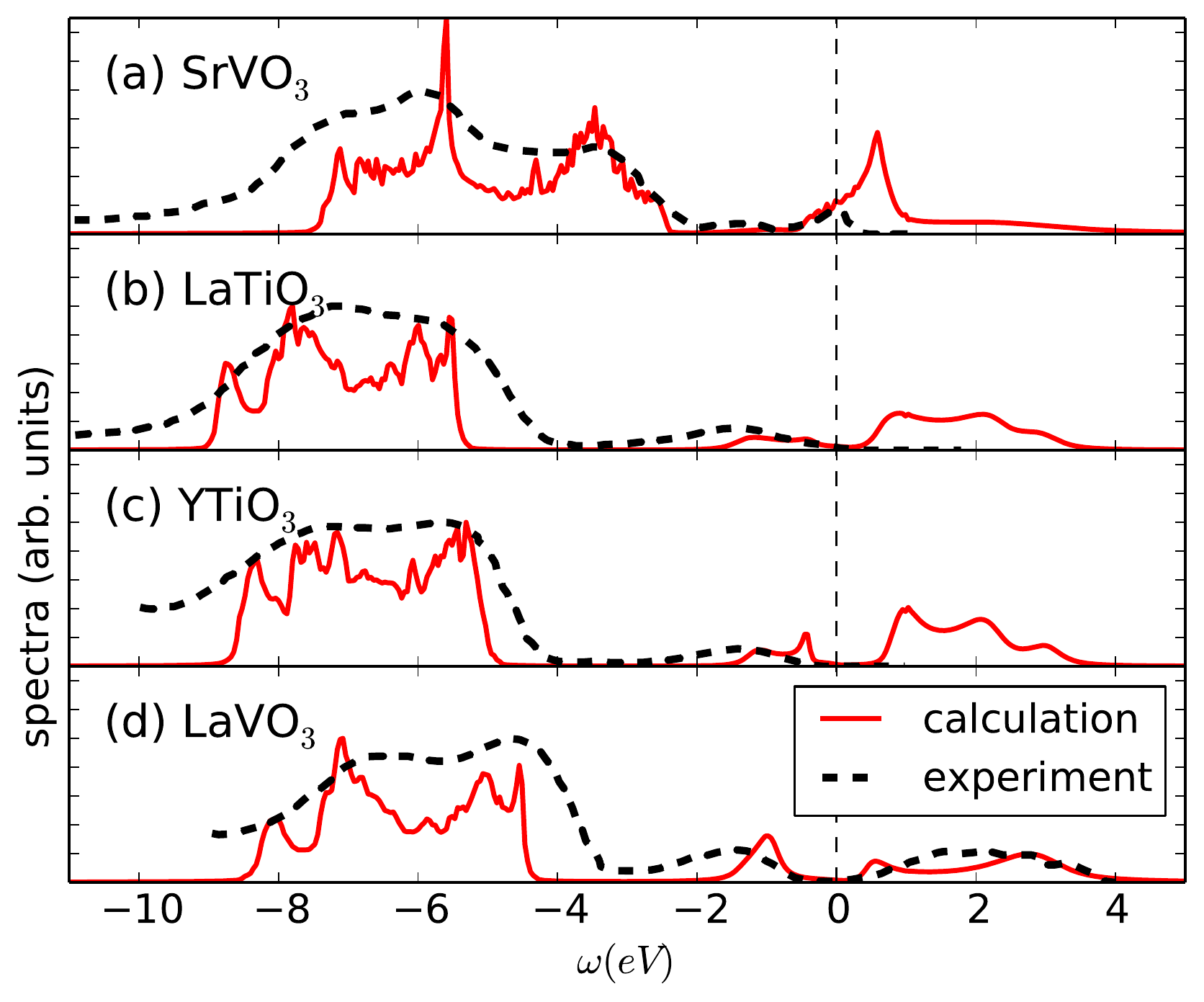}
    \caption{\label{fig:spec_with_exp}(Color online) Spectral functions $A(\omega)$ for SrVO$_3$, LaTiO$_3$, YTiO$_3$ and LaVO$_3$ using realistic lattice structure at $U=5$eV, $J=0.65$eV and $\Delta$ is adjusted to match experimental photoemission spectra (PES). The PES are from Ref.~\cite{Yoshimatsu10,Maiti00,Morikawa96,Imada98}. The vertical dashed line marks the Fermi level.}
\end{figure}

As found in Fig.~\ref{fig:dmft_la_series}, the $d$ occupancy must be reduced to drive the $d^1$ and $d^2$ systems (LaTiO$_3$, LaVO$_3$ and also implying for YTiO$_3$) into insulating state. Therefore, in one-shot DMFT, the double-counting correction must be decreased to reduce the $p$-$d$ covalency, and thus reduce $N_d$. Figure~\ref{fig:spec_with_exp} shows the spectra with the double-counting correction adjusted in order to match the experimental spectra. In this figure, with $U=5$eV, the calculated spectra are compatible with the experiments not only for the oxygen $p$ band position but also the energy gap for insulators. The results clearly show that applying  the standard FLL double counting to the computed band structure is inappropriate.

Thus to summarize, for all reasonable values of $U$, the standard scheme of FLL double counting plus the fully charge self-consistent DFT+DMFT procedure yields for LaTiO$_3$ ($d^1$) and LaVO$_3$  ($d^2$) materials (and implying for YTiO$_3$) a $d$ occupancy which is very close to that predicted by the underlying DFT calculation, and for this $d$ occupancy the materials are predicted to be metals, in contrast to experiment, which finds them to be Mott insulators. The  phase diagrams in Fig.~\ref{fig:dmft_la_series} suggest that to fix this discrepancy,  the $N_d$ value must be smaller than the DFT values and the FLL predicted values for LaTiO$_3$ and LaVO$_3$. Equivalently, the oxygen bands must lie lower in energy than predicted by the DFT calculations and the DFT+DMFT calculations which use the standard FLL double counting. The data presented in Table II makes this argument quantitative, showing the $p$-$d$ energy splitting, parametrized as the position of the oxygen bands relative to the Fermi level (defined in insulators as the middle of the gap) obtained from experiment and from DFT calculations. We see that the $p$-$d$ splitting predicted by the DFT calculations (which, as discussed above, is almost the same as the $p$-$d$ splitting obtained from fully charge self-consistent DFT+DMFT calculations with the FLL double counting) is significantly smaller than the experimental $p$-$d$ splitting. Table II also presents the $d$ occupancy obtained from DFT calculations and from one-shot calculations with the $p$ level adjusted to the experimental values. We see that the DFT calculations overestimate the $p$-$d$ covalence. We therefore suggest that one should focus on the position of oxygen $p$ bands to locate the material on the phase diagram.

\section{Comparison to  the $d$-only correlated subspace\label{sec:donly}}

Previous sections showed that $p$-$d$ covalency is important. However, unlike the case of ``late'' transition-metal oxides such as the nickelates and cuprates, materials \cite{Zaanen85} where the oxygen $p$ bands are close to the Fermi level and play an essential role in determining the physics,  the relatively large $p$-$d$ splitting characteristic of the early transition-metal oxides suggests that a $d$-only correlated subspace may capture important aspects. In this section, we will compare the two approaches. For simplicity, we will use the term ``$d$-only model'' and ``$p$-$d$ model'' to refer to the use of a correlated subspace created from frontier orbitals near the Fermi energy and well localized atomic orbitals, respectively.

For calculations with the $d$-only model, the basic DFT+DMFT framework is reapplied. The only change is in the construction of the correlated subspace: in the $d$-only model the energy window must be reduced to include only the $t_{2g}$ bands (assuming, for simplicity, that we deal  only $d^1$ and $d^2$ systems  in which $e_g$ bands do not make any significant contribution). Additionally, we will only examine the spectra of the correlated states, which means that we will not need to consider a double-counting correction and charge self-consistency will not be employed. All other steps are carried out as in the previous sections. The calculations use  the same parameters $J=0.65$eV and $\beta=10\text{eV}^{-1}$ as in the previous $p$-$d$ model calculations, while the $U$ value is reduced so that the calculated spectra have the same energy gap as in the $p$-$d$ model. The correlated subspace is defined using the MLWF method from the same DFT results used in previous section (with GdFeO$_3$-distorted structure), ensuring a fair comparison.

\begin{figure}[t]
    \centering
    \includegraphics[width=\columnwidth]{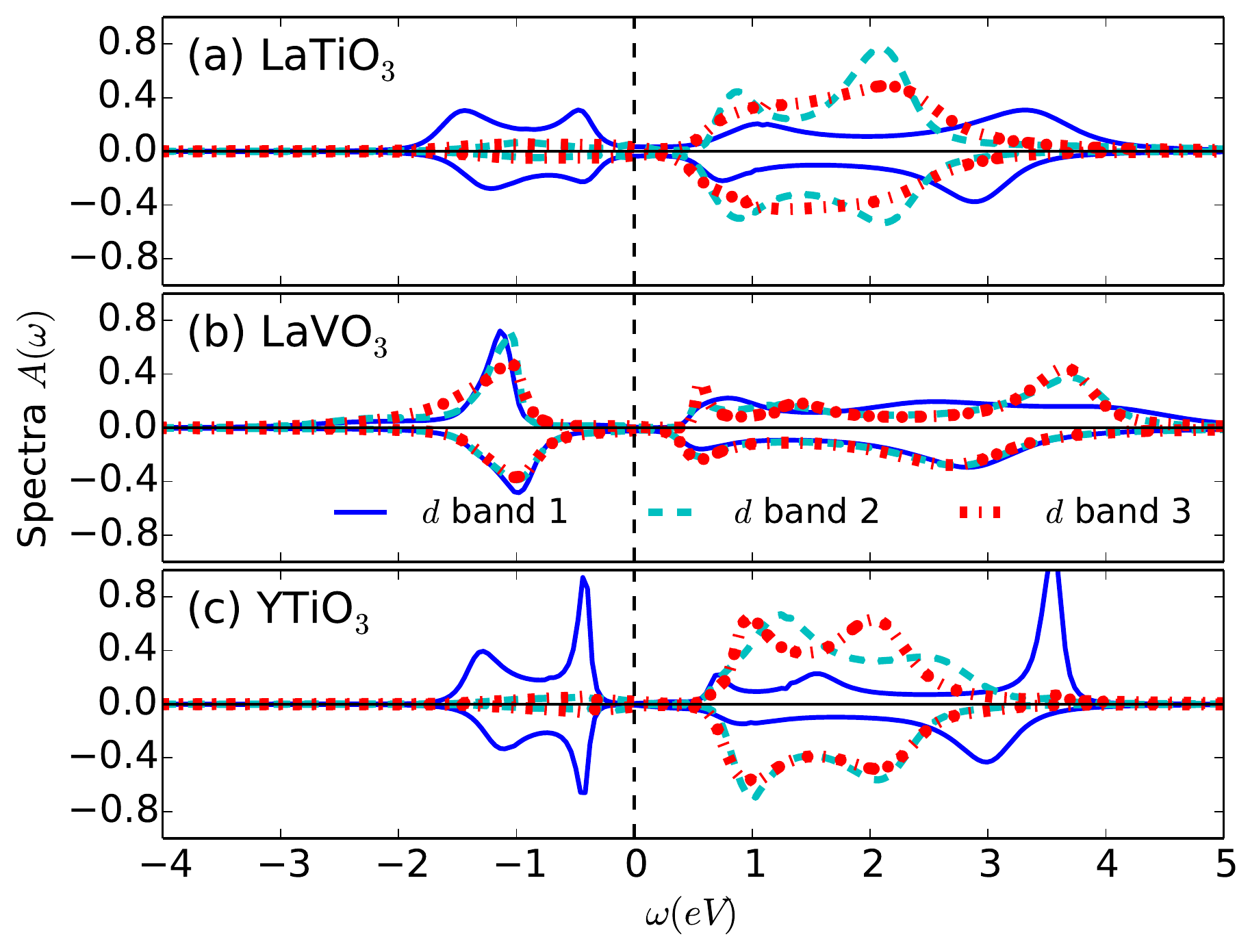}
    \caption{\label{fig:spec_pd_donly}(Color online) Comparison between $d$-only and full $p$-$d$ models for LaTiO$_3$, LaVO$_3$ and YTiO$_3$. Positive (negative) spectra are spectra of $t_{2g}$ orbitals for $d$-only (full $p$-$d$) model. The parameters $J=0.65$eV and the inverse temperature $\beta=10\text{eV}^{-1}$ are the same for both models. For the $p$-$d$ model, $U=5$eV and the double countings are set as in Fig.~\ref{fig:spec_with_exp}. For the $d$-only model, $U=4.5$eV for LaTiO$_3$ and LaVO$_3$ and $U=4$eV for YTiO$_3$. Vertical dashed line marks the Fermi level.}
\end{figure}

The  spectral functions obtained in the full ($p$-$d$) and $d$-only models  are shown in Fig.~\ref{fig:spec_pd_donly}; only the energy range relevant to the $d$-bands is displayed. Both models show the same physics: both types of spectra behave as insulator with the same orbital ordering for each materials in consideration (LaTiO$_3$ and YTiO$_3$ have ``$1$ up $2$ down'' orbital order while LaVO$_3$ has almost no orbital order). There are some differences of detail in the spectra, in particular in the positions of peaks arising  from the bands above the Fermi levels, and  the magnitude of the peaks, which are affected by $p$-$d$ covalency and subject to the uncertainties of the maximum entropy analytic continuation used here. If we define orbital order in terms of the total occupation of the $d$ level, the degree of orbital order is larger in the $d$-only model than in the $p$-$d$ model for all cases considered in Fig.~\ref{fig:spec_pd_donly}. However,  in the $p$-$d$ model some portion of the $d$-spectral weight resides relatively far below the Fermi level, at the energy of the  oxygen $p$ bands. A more reasonable comparison between the two models may be obtained by comparing the fraction occupancy of the $d$ spectrum in the energy range common to both approaches, i.e.,from $-4$eV to 0  (c.f.  Fig.~\ref{fig:spec_pd_donly}).  Considering only contributions from this energy range we find that the distribution of $d$ occupancies is, for LaTiO$_3$, $(69.7\%,14.2\%,16.1\%)$ and $(70.7\%,13.2\%,16.1\%)$ for $d$-only and $p$-$d$ models, respectively; the corresponding numbers are $(35.2\%,33.9\%,30.9\%)$ and $(35.5\%,32.9\%,31.6\%)$ for LaVO$_3$; and $(78.6\%,11.0\%,10.4\%)$ and $(79.0\%,11.3\%,9.7\%)$ for YTiO$_3$. Thus the full $p$-$d$ model is in good agreement with the $d$-only one if a reasonable effective $U$ is chosen for the latter model.

There are differences between the two models,  arising mainly from the effects of  $p$-$d$ covalency. First, in the $p$-$d$ model, there is always a $d$ portion in the bonding part of the spectra, which may cause differences in the $d$ occupancy or the orbital ordering, but these differences disappear if the same low energy window is considered for calculating the $d$ occupancy. A second effect of the oxygen $p$ bands is to reduce the electron correlation, so that to produce comparable band gaps one must use a smaller $U$ in the $d$-only model than in the full $p$-$d$ model, as shown in Fig.~\ref{fig:spec_pd_donly}.  Therefore, it appears that the $d$-only model provides a reasonable representation of the low energy physics of the  $p$-$d$ model if the interaction $U$ is appropriately renormalized. It should be noted, however, that the results presented here pertain only to the paramagnetic case. Preliminary results \cite{Dang13a}  indicate that the two models yield rather different predictions for magnetic ordering temperatures, but a full exploration of this question is beyond the range of this study.

We remark that by using the same $U$, $J$ and $\beta$ as in Refs.~\cite{Pavarini04,Pavarini05,Raychaudhury07}, we produce (not shown) very similar results for LaTiO$_3$, LaVO$_3$ and YTiO$_3$ using the same $d$-only model. While the energy gaps are similar, the orbital polarization we find is slightly weaker. In our calculations, the dominant orbital has the occupancies $0.88$ (LaTiO$_3$) and $0.91$ (YTiO$_3$), while the corresponding numbers in previous works are $0.88$ and $0.96$ \cite{Pavarini05}; for LaVO$_3$ our $t_{2g}$ occupancies are $0.73,0.68,0.59$ while Ref.~\cite{Raychaudhury07} gives $0.87,0.65,0.48$. We believe that these differences arise from differences in the construction of the $t_{2g}$ subspace. Without any correlation effect, our MLWF approach produces DOS with smaller polarization (e.g. for LaVO$_3$: $0.71,0.66,0.63$), while the method used in Ref.~\cite{Raychaudhury07} gives stronger orbital order (LaVO$_3$: $0.78, 0.63, 0.59$). Correlations will then enhance the orbital order, which explains for differences between our study and previous work, but the differences are quantitative, not qualitative. In particular we reproduce the key role played by the GdFeO$_3$ distortion which enhances the tendency to forming an   insulating state in the $d^1$ systems whereas in the $d^2$ systems, the orbital fluctuation is larger and the effect of the distortion on the insulating state is weaker. 

\section{Conclusions\label{sec:conclusions}}

In this study, we have investigated the consequences of $p$-$d$ covalency for the metal-insulator physics of  early transition-metal oxides.  We used the DFT+$U$ and DFT+DMFT  methods, with correlated subspaces defined via  projector and MLWF methods. By adjusting the $d$ level energy (i.e., the correlated subspace) and the onsite interaction, we built metal-insulator phase diagrams for materials of interest, mapping to the space of interaction $U$ and $d$ occupancy. We examined possible methods for locating materials in the phase diagrams and found that the standard FLL double-counting correction [Eq.~\ref{DeltaFLL}] gives a $d$ occupancy close to the DFT values and fails to predict the correct phase of certain materials. However, with an appropriate double-counting correction, the spectral functions match well with the experimental photoemission spectra and the metallic versus insulating nature of the predicted ground states is in agreement with experiment. We also investigated the possibility of using a correlated subspace consisting only  of delocalized, frontier orbitals (ie. $d$-only) and found that if proper parameters were used the results of well localized, atomiclike correlated subspace could be satisfactorily reproduced.

Important results obtained in this study include the following.  First, the $p$-$d$ covalency is not only important in late transition-metal oxides, as predicted by Zaanen, Sawatzky and Allen \cite{Zaanen85}, but also crucial in the early transition-metal oxides. In essence, the $p$-$d$ splitting is not larger than the important $U$ values and $p$-$d$ covalency acts to suppress electron correlation.  While we showed that effective $d$-only models can capture many aspects of the low-energy physics, for a full treatment  it is  necessary to include the oxygen $p$ bands in the calculations even for early transition-metal oxides. 

Second, the DFT+DMFT framework, with an appropriate choice of double-counting correction, gives results (photoemission spectra, energy gaps, oxygen $p$ positions) in reasonable agreement with experimental data. However, this agreement could not be obtained without experimental guidance: the double-counting correction had  to be adjusted to match with a corresponding experimental quantity (the energy gap or the oxygen $p$ band position). The  standard \textit{ab initio} methods based on double-counting corrections such as the FLL formula, in contrast, fail to put materials in the correct phase. This raises the important question of how to define a proper double-counting correction.  

Our results also confirm the importance of including realistic crystal structures. We find (as did Pavarini \textit{et al.} \cite{Pavarini04}) that the Mott insulating behavior of LaTiO$_3$ and YTiO$_3$ can only be understood in terms of the experimental (GdFeO$_3$-distorted) structure, which acts to split the $t_{2g}$ levels. 

We found that the much less computationally expensive Hartree method, and hence DFT+$U$, can well approximate certain aspects of DMFT calculations. Given a DFT+Hartree phase diagram, depending on the nominal number of $d$ electrons, one can extrapolate the DMFT phase diagram by shifting the phase boundary by an appropriate amount (see Fig.~\ref{fig:dmft_la_series}). One can get a crude picture of the DMFT paramagnetic spectra by averaging the spin up and down spectra generated by DFT+Hartree calculation. The greater computational convenience of the DFT+Hartree calculations enabled a more detailed examination of several important aspects of the physics and formalism. In particular, the DFT+Hartree calculations reveal that projector methods provide substantially more $p$-$d$ hybridization than do the Wannier methods used by many workers; this substantially affects the calculated results, and (with the different choice of double-counting correction) explains much of the difference between the results of Ref.~\cite{Haule13} and those presented here.  Understanding the origin of this difference and determining which method is more correct is an important open problem. 

We have shown that the DFT+single-site DMFT method, combined with the  phenomenological approach of adjusting the double-counting correction to place the $p$ bands at the correct  energy positions, provides a successful description of a wide range of transition-metal oxides. This suggests several  directions for future work. First, it is important to understand the evolution of the $p$-$d$ covalency  across the transition metal series as the $d$ shell is gradually filled. Extending our studies to  the materials in the crossover between early and late transition-metal oxides, in which all five $d$ bands have to be taken into account is warranted. Other aspects of the metal-insulator transition such as the temperature dependence or the metal-insulator coexistance region are also interesting topics. It is also important  to apply this model to study other properties such as spin/orbital ordering or reexamine  works done with $d$-only models to understand how the $p$-$d$ covalency affects the systems. Finally, finding an appropriate double-counting correction that correctly positions the $d$ bands relative to the oxygen bands is an important open problem. One promising approach would be to extend the $U^\prime$ ansatz \cite{Park13} to the early transition-metal oxides.

\section*{Acknowledgments}
We thank Michel Ferrero for providing an impurity solver. H.T.D. and A.J.M.  acknowledge support from  Department of Energy (DOE), Grant No. DOE FG02-04-ER046169. C.A.M. and X.A. acknowledge support from a DARPA Young Faculty Award, Grant No. D13AP00051. H.T.D. also acknowledges partial support from Vietnam Education Foundation (VEF). We acknowledge travel support from the Columbia-Sorbonne-Science-Po Ecole Polytechnique Alliance Program and thank Ecole Polytechnique (H.T.D. and A.J.M.) and J\"ulich Forschungszentrum (H.T.D.) for hospitality while portions of this work were conducted. H.T.D. acknowledges the support for computational resource from JARA-HPC. A portion of this research was also conducted at the Center for Nanophase Materials Sciences, which is sponsored at Oak Ridge National Laboratory by the Scientific User Facilities Division, Office of Basic Energy Sciences, U.S. Department of Energy, and the Extreme Science and Engineering Discovery Environment (XSEDE), which is supported by National Science Foundation Grant No. OCI-1053575.  We used the CT-HYB solver \cite{Werner06} from the TRIQS project \cite{triqs_project} as well as one written by H. Hafermann, P. Werner and E. Gull \cite{Hafermann12} based on the ALPS library \cite{ALPS1.0,ALPS2.0}.

\bibliographystyle{apsrev}
\bibliography{UNd}
\end{document}